\documentclass[nofootinbib,notitlepage,preprintnumbers]{revtex4-1}

\usepackage{amsmath,amsfonts,amssymb, amsthm}
\usepackage{mathrsfs}
\usepackage{graphicx}
\usepackage{dsfont}
\usepackage[all]{xy}

\usepackage{hyperref}
\hypersetup{
    colorlinks=true, 
    linktoc=page,    %set to all if you want both sections and subsections linked 
    linkcolor=blue,  
    urlcolor=black
}

\usepackage[usenames, dvipsnames]{color}

\newcommand{\ket}[1]{\left| #1 \right \rangle}
\newcommand{\bra}[1]{\left\langle #1 \right |}
\newcommand{\braket}[2]{\left\langle #1 | #2 \right \rangle}

\newcommand{\extd}{\mathrm{d}}

\newcommand{\hphi}{\hat{\varphi}}
\newcommand{\hphid}{\hat{\varphi}^{\dagger}}

\newcommand{\hsigma}{\hat{\sigma}}
\newcommand{\hsigmad}{\hat{\sigma}^{\dagger}}

\newcommand{\numop}{\widehat{\mathcal{N}}}

\newcommand{\move}{\widehat{\mathcal{M}}}
\newcommand{\moved}{\widehat{\mathcal{M}}^{\dagger}}

\newcommand{\allmoves}{\move_{tot}}
\newcommand{\MM}{\mathscr{M}}

\newcommand{\va}{\scriptscriptstyle}
\newcommand{\van}{\scriptstyle}

\newcommand{\ie}{\emph{i.e.}~}
\newcommand{\eg}{\emph{e.g.}~}

\newcommand{\EE}{\hat{\mathbb{E}}}

\newcommand{\iu}{\mathrm{i}}

\newcommand{\bs}{\backslash}

\newcommand{\tr}{\mathrm{Tr}}

\newcommand{\fv}{\left| 0 \right\rangle}

\newcommand{\be}{\nopagebreak[3]\begin{equation}}
\newcommand{\ee}{\end{equation}}

\newcommand{\bee}{\nopagebreak[3]\begin{equation*}}
\newcommand{\eee}{\end{equation*}}

\newcommand{\ba}{\nopagebreak[3]\begin{eqnarray}}
\newcommand{\ea}{\end{eqnarray}}

\newcommand{\baa}{\nopagebreak[3]\begin{eqnarray*}}
\newcommand{\eaa}{\end{eqnarray*}}

\newcommand{\la}{\label}
\newcommand{\n}{\nonumber}

\newcommand{\SU}{\mathrm{SU}}
\newcommand{\su}{\mathfrak{su}}

\newcommand{\suin}[2]{\langle #1, #2\rangle_{\mathfrak{su}(2)}}

\newcommand{\SO}{\mathrm{SO}}

\newcommand{\bgraph}{\mathcal{B}}
\newcommand{\vset}{\mathcal{V}}
\newcommand{\eset}{\mathcal{E}}

\newcommand{\cset}{\mathcal{C}}

\newcommand{\bint}{{\mathcal{B}_{int}}}
\newcommand{\bext}{{\mathcal{B}_{ext}}}
\newcommand{\vint}{{\mathcal{V}_{int}}}

\newcommand{\eint}{{\mathcal{E}_{int}}}
\newcommand{\eext}{{\mathcal{E}_{ext}}}
\newcommand{\einti}{\mathcal{E}_{int,i}}

\newcommand{\kernel}{J}
\newcommand{\choice}{\mathcal{C}}

\newcommand{\op}{{\widehat{\mathcal{O}}}}
\newcommand{\opkernel}{I}
\newcommand{\opgraph}{\mathcal{A}}

\newcommand{\suinter}[6]{\iota^{j_{#1}j_{#2}j_{#3}j_{#4}#5}_{#6_{#1}#6_{#2}#6_{#3}#6_{#4}}}

\begin{document}

%%%%%%%%%%%
% opening %
%%%%%%%%%%%

\title{Generalized quantum gravity condensates for homogeneous geometries and cosmology}
\author{Daniele Oriti$^1$}
\email{daniele.oriti@aei.mpg.de}

\author{Daniele Pranzetti$^2$}
\email{daniele.pranzetti@gravity.fau.de}

\author{James P.~Ryan$^1$}
\email{james.ryan@aei.mpg.de}

\author{Lorenzo Sindoni$^1$}
\email{lorenzo.sindoni@aei.mpg.de}

\affiliation{$^1$Max Planck Institute for Gravitational Physics (AEI), 
Am M\"uhlenberg 1, D-14476 Golm, Germany}

\affiliation{$^2$ Institute for Quantum Gravity,
University of Erlangen-N\"urnberg (FAU), Staudtstrasse 7 / B2, 91058 Erlangen, Germany}

\begin{abstract}
We construct a generalized class of quantum gravity condensate states, that allows the description of continuum homogeneous quantum geometries within the full theory. They are based on similar ideas already applied to extract effective cosmological dynamics from the group field theory formalism, and thus also from loop quantum gravity. However, they represent an improvement over the simplest condensates used in the literature, in that they are defined by an infinite superposition of graph-based states encoding in a  precise way the topology of the spatial manifold. The construction is based on the definition of refinement operators on spin network states, written in a second quantized language. The construction lends itself easily to be applied also to the case of spherically symmetric quantum geometries. \end{abstract}

\preprint{AEI-2015-003}

\maketitle

\tableofcontents

%%%%%%%%%%%%%%%%%%%%%%%%%%%%%%%%%%%%%%%%%%%%%%%%%%%%%%%%%%%%%%%%%%%%%%%%%%%%%%%%%%%%%%%%%%

%%%%%%%%%%%%%%%%%%%%%%%%%%%%%%%%%%%%%%%%%%%%%%%%%%%%%%%%%%%%%%%%%%%%%%%%%%%%%%%%%%%%%%%%%%

\section{Introduction}
\label{sec:Intro}
Background independent approaches to quantum gravity, in particular the strictly related loop quantum gravity \cite{LQG}, spin foam models \cite{SF}, group field theory \cite{GFT, OritiMicroDyn} and tensor models \cite{TM}, have developed greatly in recent years, with remarkable successes. They now suggest a precise type of fundamental degrees of freedom for quantum spacetime, purely combinatorial and algebraic, and define for them a well-defined fundamental dynamics, providing at the same time a plethora of mathematical tools to analyse it. Moreover, they have also produced a number of physical insights on important issues in cosmology \cite{LQC} and black hole physics \cite{ABCK, ENPP, Pranzetti}. This has been made possible by the clever application of tools and ideas from the fundamental theory to simplified models.   

What is still missing, however, is a direct and clear route from the fundamental background independent kinematics and dynamics of the theory to effective continuum geometry and spacetime-based physics. From the point of view of the formalism, this means studying the renormalisation flow of the quantum dynamics and the collective behaviour of the fundamental degrees of freedom of quantum spacetime (see, for instance, the recent developments in
the spinfoam \cite{biancarenorm} and in the GFT \cite{GFTrenorm} contexts).
From a more physical perspective, this is the problem of identifying suitable quantum states {\it within the fundamental theory} that admit an interpretation in terms of continuum spacetimes and geometries, and of extracting their effective (classical and quantum) dynamics. Notice that this {\it problem of the continuum} is distinct, even if related to the problem of studying the semi-classical limit of the same theory (for which see \cite{coherent states, OPS, asymptSF}). In particular, addressing these two issues should allow to make direct contact, again from within the fundamental theory, with the simplified models used in (quantum) cosmology and (quantum) black hole physics and to give a more fundamental justification to their many results.

It is natural to focus first on homogeneous geometries, as 
they are the simplest case to consider while still playing a key role in concrete physical situations,
for instance in cosmology. They are also the natural setting where to look for phenomenological consequences of a quantum gravity theory (see, for instance, the recent work in the canonical framework \cite{AlesciCianfrani} and in the covariant one \cite{SFcosmology}).
Among these, the line of research that we develop here is that of {\it extracting cosmology from group field theory (GFT) condensate states}, that was started in \cite{cosmoshort,cosmolong}, and further developed in \cite{Gielen, GO, fidelity}. We will discuss the main ideas of this framework in the following sections. It is based on the hypothesis that the geometric phase of group field theory, for the models that encode the quantum geometry of loop quantum gravity (LQG), is a condensate one, and that simple GFT condensate states capture the physics of homogeneous geometries. Indeed, it was shown that the effective dynamics of such GFT states, extracted directly from the fundamental one, is a non-linear extension of quantum cosmology. 

The objective of this paper is to generalise the construction of states performed in \cite{cosmoshort,cosmolong} and to introduce, in the same GFT framework,  a new class of quantum states implementing the notion of homogeneity of the associated geometry (in the same coarse grained sense). 
Among their new features, an important one is that these new states contain the basic connectivity information required to reconstruct the topological structure of the desired geometry.
While well suited for homogeneous spaces, these states can also be adapted immediately to the case of spherically symmetric geometries, where (classically) the isometry group leads to a foliation of the space in terms of homogeneous leaves. This is a very important point of our construction. Indeed, the generalisation to spherically symmetric quantum geometries will be the subject of future work based on our present results. 

More generally, our construction clarifies the construction of coarse grained continuum geometric states, with given symmetry properties, in GFT/LQG, keeping the topology encoded in such states under control and 
retaining the ``sum over triangulations'' spirit of GFT (thus not relying on any fixed lattice structure). Most importantly, our states correspond to {\it continuum quantum geometries} in the sense that they encode  (in terms of a few collective variables)  an infinite superposition involving an infinite number of spin network degrees of freedom. In the process, we show by our construction another general point: the Fock structure behind (the reformulation of LQG as) GFT allows a straightforward definition of refinement operators for spin network states, and their action connecting quantum states based on different graphs. Our specific construction is tied to the notion of ``homogeneity" (in our present scheme, this translates into ``wave-function homogeneity'', as we shall discuss) and this will impose peculiar and stringent properties on the refinement operators we use. However, the general idea of constructing superposition of spin network states involving arbitrarily refined graphs, and thus being candidates for continuum quantum spaces, via refinement operators defined in terms of GFT field operators, is very general. We are confident that it will find a wide range of applicability and lead to further progress in the future.

Let us stress that our results can also be seen as an LQG construction, as the GFT re-formulation \cite{GFT-LQG} carries the same type of data and combinatorial structures as canonical LQG, even if it encodes them in a different Hilbert space.

Also, in the present work some tools developed for tensor models and already heavily used in the GFT context, show directly their potential usefulness in a more physical and LQG-related context. This is not surprising, given that these tools were developed to have a better control over (superpositions of) combinatorial structures, and this control is crucial for our construction of continuum states of homogeneous type. 
Still,  this is one more development that is likely to lead to further progress in the future.

We start in Section \ref{sec:overview} with a brief overview of the GFT Fock space.
In Section \ref{sec:GenPrin} we outline the general principles to guide us to a GFT realization of homogeneous geometry, clarifying how we seek to generalise the construction in \cite{cosmoshort,cosmolong}. 
In Section \ref{sec:operators} we introduce the field operators which represent the basic tools of our construction and apply them in a simplicial set-up.
Section \ref{sec:GraphTop} presents some definitions of the main graph structures encoding topological information. 
 Section \ref{sec:melonic} contains an extension of the tools developed in the simplicial context to colored graphs and melonic structures. 
In Sections \ref{sec:sphere} and  \ref{sec:shells} we apply the general techniques to some specific topologies: we show how to construct (considerably) arbitrary triangulations of a 3-sphere and a shell at a given radial coordinate respectively. 
Section \ref{sec:vev} analyzes the expectation value of generic one-body operators and contains an explicit calculation for the number operator.
Section \ref{sec:geometry} presents the second quantization of LQG geometric operators allowing us to extract geometric information out of these states.  Section \ref{sec:conclusions} contains the conclusions and some final remarks.  We included some auxiliary technical material in the Appendix A.

\section{GFT in the Fock space picture: a brief overview}
\la{sec:overview}

In a nutshell, Group Field Theories (GFTs) are quantum field theories on group manifolds, defining a path integral quantisation of the gravitational field. To achieve this objective, one builds the theory so that its perturbative expansion yields Feynman diagrams interpretable as discrete geometries. In principle, GFTs can model any chosen spacetime dimension, while the amplitudes are determined by the selected discrete counterpart of gravitational/spacetime dynamics \cite{OritiMicroDyn}.

The GFT program can be seen as a generalisation to higher dimensions of the matrix model approach to two-dimensional quantum gravity \cite{matrixmodels}. Moreover, while they share with tensor models \cite{TM} the same combinatorial structures in both field kinematics and dynamics, at the level of action and Feynman diagrams, GFTs are also an extension in that they weight the Feynman diagrams by additional data. These data are determined by the group manifold on which the group fields are defined. This allows a comparison (or a specific relationship) with other approaches to quantum gravity. In particular, one can easily construct models whose perturbative expansion gives the same amplitudes of any given spin foam model, thus incorporating fully the latter framework.
For an extensive discussion of these points, see \cite{OritiMicroDyn}. 

For the purposes of this paper, it is important to stress another aspect of GFTs: they constitute a translation, in the language of second quantisation, of Loop Quantum Gravity. This is most easily seen when GFT is written in its operatorial version, in which field operators act on a Fock space (see \cite{GFT-LQG} for a detailed presentation). This formulation allows a direct description in terms of spin network states and quantum operators acting on them.

To avoid confusion, we consider GFT models for gravity in 3+1 dimensions. 
To maintain clarity, we examine the simplest version of the theory. Generalisations are possible and straightforward, but they might obscure the relevant ideas. 
In this context, the basic components of the theory are the \emph{group field creation/annihilation operators}: $\hphid$ and $\hphi$, respectively.  
The operators possess\footnote{This symmetry property, which endows the Feynman amplitudes with a lattice gauge theory structure, could in principle also be implemented at the dynamical level, rather than in the very definition of fields.} a \emph{gauge invariance property}, symmetry under the diagonal right action of $SU(2)$
on $SU(2)^{\times 4}$:
\begin{equation}
  \la{gauge-inv}
  \hphi(g_vk) = \hphi(g_v)\;,
\end{equation}
where $k\in\SU(2)$ and, to clarify notation, $g_vk \equiv (g_{(v,1)}k,g_{(v,2)}k,g_{(v,3)}k,g_{(v,4)}k)$.

As a consequence of this right gauge invariance, the operators satisfy \emph{bosonic commutation relations}:
\begin{equation}
	\label{eq:BosComRel}
	[\hphi(g_v), \hphid(g_{v'})] = \Delta_{R}(g_v, g_{v'}) \equiv \int_{\SU(2)} d\gamma 
	\prod_{i=1}^4 \delta(g_{(v,i)} \gamma g_{(v',i)}^{-1})
\end{equation}
where  $g_v \equiv (g_{(v,1)},g_{(v,2)},g_{(v,3)},g_{(v,4)})$ and $\delta(g)$ denotes the $\delta$-function over $\SU(2)$. Note that this is akin to the Fock structure of a non-relativistic QFT.

These ladder operators have a graph-theoretic manifestation as operators creating or destroying 4-valent vertices. This is enhanced to a topological interpretation when such vertices are thought of as dual to tetrahedra. The gluing of 4-valent vertices constructs 4-regular graphs, which is mirrored in the dual by the gluing of tetrahedra to construct three--dimensional simplicial topologies. As a result, all Fock space states are naturally associated to graphs and, thereafter, $3d$ topologies, whose properties depend on the contraction of group arguments within the product of operators. 

A local geometric structure is captured precisely by the data associated by the group field operators to each tetrahedron; the group elements are interpreted directly as a discrete connection (that is, a collection of group elements, which, in any embedding of the combinatorial structures into a smooth manifold, acquire the interpretation of parallel transports of a spacetime connection along paths in the manifold), parallel transporting information from (the frame attached to) one tetrahedron to its neighbour (when glued, otherwise it is the local value of the discrete connection).  The geometrical interpretation is more clearly elucidated if we pass to the so-called non-commutative representation \cite{GFTnoncomm,Baratin}. This is 
obtained by performing a (non-commutative) Fourier transformation on the field operators, passing from the group $\SU(2)$ to the corresponding algebra $\su(2)$.  These \emph{algebra field operators} are denoted by $\hat{\phi}$ and $\hat{\phi}^\dagger$, respectively.  The domain of these fields is $\su(2)^{\times 4}$ and the transformation takes the form:   
\begin{equation}
  \hat{\phi}(X_v) \equiv \int \extd g_v \left[\prod_{i=1}^{4}e_{g_{(v,i)}}(X_{(v,i)})\right]\hphi(g_v),
\end{equation}
where the Haar measure $\extd g_v\equiv\prod_i dg_{(v,i)}$, $e_{g_{(v,i)}}(X_{(v,i)})$ are the analogue of plane waves in this non-commutative context ($e_{g}(X) = e^{\tr_{\su(2)}(Xg)}$).
This quartet of algebra elements captures discrete metric information for the tetrahedron, to which the field operator is associated.  In fact, the four Lie algebra elements associated to each tetrahedron can be interpreted as defining the normal vectors to its four faces, and can be used to define a local tetrad frame. $|X_{(v,i)}| = \sqrt{\suin{X_{(v,i)}}{X_{(v,i)}}}$ and $X_{(v,i)}/|X_{(v,i)}|$ denote then the area of and the normal to the $i$th triangle, respectively.  
In this formulation, the gauge invariance condition translates into the \emph{closure condition}:
\begin{equation}
  \sum_i X_{(v,i)}=0\,.
\end{equation}
Having imposed closure, one can construct six SU(2)-invariant quantities that capture the remaining information, for example,  the six tetrahedral edge-lengths.  Thus, the connection between wave-function data and local geometric structure follows.

It is worth noting that these two representations, $g$ and $X$, are akin to the usual configuration space and momentum space representations in QFT. In the classical regime, the two together parameterise the discrete geometry phase space, which matches exactly the classical phase space associated to the corresponding graph in LQG.
Obviously, other parametrizations of the same phase space are possible \cite{twistor}, and give rise to corresponding representations for the GFT fields. 

A generic $n$-particle Fock state is built from creation operators acting on the Fock vacuum:
\begin{equation}
  \label{eq:GenWave}
  \ket{\psi} = \int \prod_{n = 1}^N \extd g_{v_n}\; \psi(g_{v_1},\dots,g_{v_N}) \,
  \hphid(g_{v_1})
  \ldots
  \hphid(g_{v_N})
  \fv ,
\end{equation}
where  $\fv$ denotes the Fock vacuum, $\hphi(g_v)\fv = 0$, and $\psi$ is an $n$-body wave-function. Notice that the latter is completely symmetric under permutation of its arguments, due to the bosonic statistics satisfied by the field operators.

The wave-function has a topological interpretation in so far as it determines the topological properties of the collection of $n$ tetrahedra comprising the quantum state; the function $\psi$ encodes the convolution of group field arguments. This translates to connectivity among the tetrahedra.  Such wave-functions have an immediate interpretation as spin network wave-functions within LQG. 
From the point of view of LQG, the field theory ladder operators connect spin network wave-functions with different number of nodes.

The dynamics of the theory are determined by the \emph{quantum equation of motion}, which generically has the following functional form: 
\begin{equation}
  \frac{\delta\mathcal{S}}{\delta\overline{\varphi}(g_v)}[\hphi,\hphid]\ket{\Psi}  = \left(\int \extd g_{v'}\; \mathcal{K}(g_v,g_{v'})\, \hphi(g_{v'}) + \frac{\delta \mathcal{V}}{\delta \overline{\varphi}(g_v)}[\hphi,\hphid] \right) \ket{\Psi}= 0
\label{GFTEOM}
\end{equation}
where $\mathcal{S}[\varphi,\overline{\varphi}]$ is the action functional,\footnote{Note that the action is a functional of \emph{group fields} ($\varphi$ and $\overline{\varphi}$) at the outset and not group field operators ($\hphi$ and $\hphid$); hence, the absence of \lq\lq hats\rq\rq.  After the functional derivative is taken, the operators are substituted into the result. This requires a choice of operator ordering, and normal ordering is the obvious choice. In general, thanks to the GFT formulation, one can follow standard QFT procedures, even in this background independent quantum gravity context.} $\mathcal{K}$ is a kinetic kernel encoding the free propagation of the field, while
$\mathcal{V}[\varphi,\overline{\varphi}]$ is the vertex term, which encodes the interacting part of the theory.
In particular, it contains (non-local) non-linear monomials in the field operators. 
The specific choice of the equation of motion allows us to translate the LQG dynamics to GFT form. The action can indeed be chosen to encode the quantum dynamics of a canonical theory of spin networks, like LQG, via projector operator onto solutions of the Hamiltonian constraint, see \cite{GFT-LQG}.

The perturbative expansion is defined in the usual way, starting from the state that solves
the free theory equation of motion, \ie the Fock vacuum
\begin{equation}
	\label{eq:Free}
	\int \extd g_{v'} \;\mathcal{K}(g_v,g_{v'})\,\hphi(g_{v'})\;\fv = 0
\end{equation}where we implicitly assume that $\mathcal{K}$ is nondegenerate.
With an appropriate choice for the vertex kernel, a generic
term in the expansion is associated to spin network states, along with a complex coefficient determined by \eqref{GFTEOM}. In this way, this operatorial approach to quantum gravity also generates a sum-over-geometries.

The Fock structure facilitates working with solutions to \eqref{GFTEOM} that are comprised of 
arbitrary superpositions of states containing an arbitrary
numbers of \lq\lq quanta of space\rq\rq. These are candidates for continuum states of quantum geometry in the theory, as they do not depend on any given combinatorial structure nor on any given truncation to a finite number of degrees of freedom. Furthermore, the quantum equations are generically too complicated to be solved explicitly, or, even in those cases for which it is possible to write an exact expression for the state, its physical content remains implicit. Thus, the Fock structure also provides 
the possibility to utilise completely new classes of trial states as 
approximate solutions, possibly with a clearer geometric content. 

In particular in \cite{cosmoshort,cosmolong} (see also \cite{Gielen, GO}), powerful Fock space tools (coherent states, squeezed states and generalizations)
allowed for the derivation of effective dynamical equations for states that admit a natural interpretation
as homogeneous cosmologies. Importantly, this was accomplished while still preserving the sum-over-geometries character of
the theory, in that no preferred triangulation was singled out as an identifiable background.

The rest of this paper is devoted to the application of these ideas and methods to the construction
of states that can be interpreted in a natural way as superpositions of spin networks states, associated to complexes of given topology, and similarly interpreted as encoding in a coarse-grained manner the quantum geometry of continuum homogeneous spaces.

%%%%%%%%%%%%%%%%%%%%%%%%%%%%%%%%%%%%%%%%%%%%%%%%%%%%%%%%%%%%%%%%%%%%%%%%%%%%%%%%%%%%%%

%%%%%%%%%%%%%%%%%%%%%%%%%%%%%%%%%%%%%%%%%%%%%%%%%%%%%%%%%%%%%%%%%%%%%%%%%%%%%%%%%%%%%%

\section{General principles guiding the construction of homogeneous quantum geometries}
\label{sec:GenPrin}
Let us now present in some detail the general philosophy guiding our approach.

Our general desiderata are the following. First, we want to identify quantum states {\it in the full theory}, which a) contain an infinite number of kinematical degrees of freedom; b) do not depend on any given triangulation/graph but are generically a superposition of (possibly an infinite number) of them; c) depend only on data associated to continuum homogeneous (quantum) geometries.

This means that their description of quantum geometry is necessarily a {\it coarse-grained} one, with respect to the infinite number of degrees of freedom associated to their constituent microscopic configurations (the states appearing in their defining superposition). This also means that we are {\it not} requiring necessarily that their constituent microscopic configurations are themselves homogeneous discrete quantum geometries, \ie states associated to discretizations of homogeneous geometries,  although these latter would fit the general desiderata as included in the superposition. We will return later to this point.

Second, we would like to identify states that are also candidates for good approximations to the geometric vacuum of the full theory, that is the vacuum state in the phase which admits a description in terms of smooth metric fields and (some modified version of) General Relativity (this phase should of course exist, barred the failure of the theory as a theory of quantum gravity). Here, the key word is ``approximate". The goal is to find reasonable simplified trial states, that capture enough of the true vacuum of the system to allow the extraction of accurate physical predictions. The hypothesis is that a first approximation of such true vacuum would be described in terms of homogeneous geometric degrees of freedom, suitable for the description of (quantum) cosmology.

Our next guideline is then the idea of condensation of the atoms of quantum space as the process leading to the geometric phase of the theory, i.e. the idea of the universe as a quantum condensate of QG building blocks (in this context, GFT/LQG building blocks) \cite{Hu}. Such ``GFT condensation" is also a prototypical example of the phase transitions that we can expect in these quantum gravity models \cite{danieleemergence}. Indeed, its occurrence in GFT has been studied using FRG methods in \cite{FRG-GFT}, and suggested in the canonical LQG context in \cite{Tim, Hanno-Tim}. The idea of condensation leads immediately to a specific guess for the quantum states to consider (again, we stress, as simple enough approximations of a true vacuum of the theory). Under the assumption of bosonic statistics for the GFT quanta, a GFT condensate would be a quantum state in which all quanta appearing in its Fock representation are associated the same wave function. Now this is a strong constraint on the class of quantum states to consider, and we will stick to this restricted class in the following.

The simplest example of states of this kind has been considered in \cite{cosmoshort,cosmolong}, and shown to admit indeed an interpretation in terms of homogeneous (but generically anisotropic) continuum geometries, and to lead to an effective dynamics that is a non-linear generalization of loop quantum cosmology. 

The simplest example of a GFT condensate state is a coherent state for the field operator:
\be |\sigma\rangle := \mathcal{N}(\sigma) \exp\left(\hat\sigma\right)|0\rangle \quad\text{with}\quad \hat\sigma := \int \extd g_v\; \sigma(g_v)\hat\varphi^{\dagger}(g_v) \,
\ee
where we also require $\sigma(k g_v )=\sigma(g_v)\, \forall k\in SU(2)$, and $\mathcal{N}(\sigma)$
is a normalization factor. 

In series expansion, one obtains an infinite superposition of $n$-particle states of n disconnected tetrahedra (equivalently, spin network vertices). One has something akin to a gas of disconnected tetrahedra, 
where all $n$-particle wavefunctions have a product structure built out of a single wavefunction $\sigma$ depending on the data of an individual tetrahedron. 
This wavefunction $\sigma$ then plays the role of a collective wavefunction characterising the whole GFT condensate state, that is, the whole set of infinite degrees of freedom encoded in such coherent state. Indeed, this is the exact GFT counterpart of the Gross--Pitaevskii wave function of standard BECs \cite{leggett}.

One can show that the phase space on which this collective wave-function is defined, thanks also to the additional symmetry condition we imposed, can be put in one-to-one correspondence with the data characterizing homogeneous continuum spatial geometries. For details, see \cite{cosmoshort,cosmolong}.

This correspondence relies on a choice of embedding of the abstract tetrahedra in a continuum 3-manifold, and a corresponding re-interpretation of the abstract algebraic data defining the single-tetrahedron phase space as coming from the discretization of smooth fields defined on this manifold.  It is in this sense that they are good candidates for describing quantum homogeneous geometries in the full theory. In this embedding, each tetrahedron encodes the local structure of geometry at each point, and the equality of the associated wave functions is the translation of the requirement of homogeneity. 

The choice of such embedding is not intrinsic to the formalism, it is instead an (ambiguous) extra ingredient needed for the interpretation of the data and of the states, but nothing in the construction itself depends on it. 
The resulting dynamics that that should confirm (or challenge) the interpretation of the states based on such choice of embedding. 

In fact, one way to see the construction is as {\it defining} a notion of homogeneity in a fully background independent context, a notion that we may call {\it wave-function homogeneity}: a quantum geometry is homogeneous if the (quantum statistical) distribution over pre-geometric data associated to each of the quanta of space is the same. 

The coarse grained nature of the above states rests on the fact that the infinite superposition of spin network degrees of freedom is fully encoded in a single collective function of only a finite number of variables. This is according to our desiderata for cosmological states in the full quantum gravity theory.

However, no correlation at all between GFT quanta (tetrahedra, spin net vertices) 
is encoded. The latter condition means also that these state lack the information to reconstruct any topological structure.

They are thus only a crudest possible approximation to any realistic vacuum state of the interacting theory (they correspond, indeed, to a simple mean field approximation), even assuming condensation correctly captures the main features of its geometric phase. 

We seek for a generalization of the above construction, and for more realistic 
states satisfying our desiderata.

Any deviation from the simplest structure of the GFT coherent states will introduce new data in the states entering the superposition. 
Now, each individual state entering the superposition will correspond, generically, to an inhomogeneous discrete (simplicial) geometry. \footnote{Note that this is true saved a few exceptions, because the same states may include also, in their defining superposition, states associated with triangulations of regular combinatorial structure as well as homogeneous assignment of discrete data. We will return to this point.}

We look now for states that, while defined by infinite superpositions involving triangulations and discrete quantum geometries which are not individually homogeneous, they are still collectively described by the data associated to continuum homogeneous geometries, thus still matching our 
desiderata 

Generically, modifying the wave function assignment will lead to a large number of variables being needed for characterizing the quantum state. States of this type are of course acceptable, but very impractical if the goal is to insert them in the fundamental dynamics of the theory to extract an effective dynamics. If, in order to avoid this, one simply uses a single wave function on a restricted set of variables to collectively characterize the whole superposition, \eg by simply integrating out all the other microscopic data (a brute force coarse graining), one should expect the effective dynamics satisfied by this wave function to be vastly different from the microscopic one defining the full theory. This is not conceptually problematic, but it makes technically very hard to extract such effective dynamics from the microscopic theory.

We stick, then, to what we called {\it wave-function homogeneity} as a defining property, for now. Therefore, the states are still going to be characterized collectively by a single function on minisuperspace, of (quantum) cosmological type.
However, they also encode correlations among such building blocks that allow to associate one and the same topology to each term in their defining superposition\footnote{Notice that this rests on the peculiarity of GFTs as combinatorially non-local field theories, i.e. on their richer combinatorial structures. In ordinary QFT, like the one describing standard BECs, once the assumption of identical wave functions for the microscopic constituents is implemented, there is not much room for any other condition on the states.}. Here lies their main improvement over the simplest examples of GFT condensates. As topological information is encoded in the graph structure, we shall examine this in some detail in Section \ref{sec:GraphTop}.  In this paper, we shall treat three cases, the 3-sphere, the 3-ball and the 3-shell. They are compact manifolds with zero, one and two boundary components, respectively. The first is of interest in its own right, while we foresee the latter two as important for the case of spherical symmetry.

In more detail, we explicitly realise our states via a two-step process:
    \begin{itemize}
      \item[i)] we begin with a \emph{seed state} for the desired topology.  A seed state is generally chosen for it combinatorial simplicity.
      \item[ii)] we act iteratively with a \emph{refinement operator}, that generates states associated to finer triangulations from simpler ones.  In particular, this operator is both topology-preserving and maintains homogeneity of the vertex wave-functions. 
    \end{itemize}
    
The states obtained by this process are a linear superposition of states with support on structures drawn from a subset of triangulations of that topology. Considering a single contribution to that sum, the expectation value for observables probing geometric structure is determined exclusively by the unique vertex wave-function and the combinatorics of the graph.  Once more, this means we are capturing only coarse grained information about the multitude of microscopic degrees of freedom that can be associated to the individual states in the superposition. In particular, like in the case of simple condensate states, one can see that this coarse grained information matches the one characterising the geometry of homogeneous spaces, as they are used in cosmology.

One could summarise this by saying that we plan to have homogeneous wave-functions over discrete geometries, rather than wave-functions over homogeneous discrete geometries.  

Reproducing 
a classical homogeneous geometry 
in a discrete context is a subtle procedure,   
as, in the discrete setting, there are extended as well as point-like structures: points, edges, faces, 3--cells, all of which have subtly different realisations of locality. 

This becomes evident when we assign geometric information. For example, one familiar option is to assign metric degrees of freedom within a frame attached to each tetrahedron,
as it is done in the GFT non-commutative $X$-representation.  Thus, discrete metric information is local to each tetrahedron.  However, curvature information
cannot be local to a given tetrahedron, since it describes precisely the effect of transporting between frames. 

With this in mind, one looks for a suitable definition of homogeneous simplicial geometries.  

One option focusses on assigning homogeneous geometrical information to the triangulation, leaving implicit the definition of the transitively acting isometry group. While metric information poses no problem,  
curvature information is a different matter. Since curvature is local to the edges, homogeneity would impose regularity of their neighbourhoods. Triangulations satisfying both requirements are very special and difficult to locate within the space of GFT triangulations (at least in dimensions three and higher). The same is true in the even larger space of complexes associated to generic LQG states. As a result, they are likely to have little impact within the GFT setting and any quantum state relying exclusively on a superposition of these very special complexes would probably be, maybe paradoxically, an even worse approximation to the true vacuum of the theory (still assuming this corresponds to homogeneous quantum geometries) than the simplest condensate states. 

In light of this fact,  another option would be to stress more the role of an isometry group of transformations.  In this way, we could increase the number of relevant triangulations by relaxing the combinatorial regularity conditions \emph{and} compensating for this irregularity with varying geometrical information assigned to the discrete structures. In other words, the geometrical degrees of freedom could no longer be homogeneous, in the sense that they would vary over the tetrahedra of the triangulation.   As a result, this also scuppers our plans from a GFT perspective, since it entails a proliferation of elements within the fundamental data set.  
This conflicts with one of the main motivations for pursuing a homogeneous sector within GFT: ease of construction.  

Thus, quantizing a classically homogeneous sector of discrete geometries does not seem well-adapted to the GFT setting, and an approach rather of coarse graining type seems more appropriate.  

In the end, the states constructed by our approach possess three important properties:
\begin{itemize}
  \item[--] they are reasonably simple to generate and manipulate, providing the opportunity for explicit analytic calculations;
  \item[--] they are compatible with a statistical mechanical approach that takes into account a large class of microstates at fixed macroscopic conditions; 
  \item[--] they describe a macroscopic continuum geometry controlled by the minimal amount of data: an infinite but well-controlled class of graphs describing a fixed topology, along with a single vertex wave-function.
\end{itemize}

Having clarified the conceptual set-up and our goals, we can now move on to the technical part, and describe our construction of generalised quantum gravity condensate states in detail.

%%%%%%%%%%%%%%%%%%%%%%%%%%%%%%%%%%%%%%%%%%%%%%%%%%%%%%%%%%%%%%%%%%%%%%%%%%%%%%%%%%%%%%

%%%%%%%%%%%%%%%%%%%%%%%%%%%%%%%%%%%%%%%%%%%%%%%%%%%%%%%%%%%%%%%%%%%%%%%%%%%%%%%%%%%%%%

\section{Redefinition of field operators}%: vertex homogeneity and refinement moves}
\la{sec:operators}

We have two goals here: to realise homogenity of the vertex wave-function 
and to define refinement moves for the topologies of interest. In this section we will consider just the case of GFT without matter and without additional colouring. The extension of the results presented in the following is
straightforward.

To achieve our goals, we  have  to adapt the formalism in order
to introduce operators which create vertices with the desired wave-function.

\subsection{Vertex wave-function homogeneity}\la{sec:vertex-WF}

It is convenient to define field operators, functions of the original GFT field operators  $\hphi$ and $\hphid$,
which are automatically adapted to the wave-function that we want to encode. More precisely,
we are interested in linear transformations of the form\footnote{Generically, one may perform linear transformations of the form:
\begin{eqnarray}
	\hsigma(g_v) &=& \int \extd h_v \Big[ \alpha(g_v; h_v)\,\hphi(h_v) + \beta(g_v;h_v)\,\hphid(h_v)\Big]\;,\\
\hsigmad(g_v) &=& \int \extd h_v \Big[\overline{\alpha}(g_v; h_v)\,\hphid(h_v) + \overline{\beta}(g_v;h_v)\,\hphi(h_v)\Big]\;,
	\end{eqnarray}
	The effect on the commutation relations of such a transformation is:
	\begin{equation}
	  [\hsigma(g_v), \hsigmad(g_{v'})] = \int \extd h_v\Big[ \alpha(g_v; h_v)\,\overline{\alpha}(g_{v'};h_v) + \beta(g_v; h_v)\,\overline{\beta}(g_{v'};h_v) \Big]\;.
	\end{equation}
	Thus, in order to be a Bogoliubov transformation, one should impose the condition:
	\begin{equation}
	  \int \extd h_v\Big[ \alpha(g_v; h_v)\,\overline{\alpha}(g_{v'};h_v) + \beta(g_v; h_v)\,\overline{\beta}(g_{v'};h_v) \Big] = \delta(g_v, g_{v'})\;.
\end{equation}
In Equation \eqref{field}, we are considering a transformation, in which $\alpha(g_v;h_v) = \sigma(g_vh_v^{-1})$ and $\beta(g_v;h_v) = 0$.  }
\begin{equation}
	\label{field}
	\hsigma(h_v) = \int \extd g_v\; \sigma(h_vg_v) \,\hphi(g_v)\,, \quad\quad \hsigmad(h_v) = \int \extd g_v\; \overline{\sigma}(h_vg_v) \,\hphid(g_v) \,,
\end{equation}
where  $\hsigma$ and $\hsigmad$ are the transformed fields. 
Let us comment for a second on the interpretation of the group elements appearing in the previous
expressions. The group elements $g_v$ are those associated to a vertex, and they can
be seen as the parallel transports from the vertex to the endpoint of an open edge emanating from
it. 

The group elements $h_v$ do not possess this geometrical interpretation. Rather, they are
auxiliary group elements used to manipulate the connectivity encoded in the state. Indeed,
the connectivity is encoded in the functional dependence of the wave-function of the state with
respect to the parallel transports $g_v$. We say that there exist an edge connecting two vertices,
$v$ and $w$, if the wave-function depends on $g_v$ and $g_w$ through the specific combinations
of arguments $g_{(w,j)}^{-1}g_{(v,i)}$ for some pair of indices $(i,j)$. 

A simple way to generate
these gluings, starting from wave-functions associated to the vertices, is through convolutions. For instance, an edge built using the leg $1$ of vertex $v$ and the leg $3$ of vertex $w$ will be built
with the following convolution
\begin{equation}
\psi_{v}(g_{v})\psi_{w}(g_{w}) 
\rightarrow 
\int dh\psi_{v}(hg_{(v,1)},g_{(v,2)},g_{(v,3)},g_{(v,4)})\psi_{w}(g_{(w,1)},g_{(w,2)},hg_{(w,3)},g_{(w,4)}) .
\end{equation}
Notice that the position of the group element $h$ on the left of the $g$s is dictated by
 the requirement that the resulting wave-function will depend on the group element $g_{(w,3)}^{-1}g_{(v,1)}$, which is the parallel transport from $v$ to $w$ along the desired path. 

Therefore, the role of the group elements $h$ appearing as the arguments of the operators
$\hsigma$ is to be used as the auxiliary variables required by the convolutions necessary
to encode the connectivity of the states.

In the case in which the desired state is associated to an open graph, it suffices to act with the operator $\hsigmad$ evaluated at
$h_i = e$, where $i$ is the label of the argument that corresponds to the open edge.

At this stage we introduce a further restriction on the wave-functions $\sigma$, which can be
motivated by geometric considerations. The wave-functions that we are using are encoding
the geometric data of a tetrahedron. However, in general, they store more information that
we need, if we just impose gauge invariance on the right, as they specify, in terms of the 
Lie algebra variables, not only the edge lengths, but also the orientation of the tetrahedron
itself. However, as it has been argued in \cite{cosmoshort,cosmolong}, this is too much, as it does not reflect
the gauge invariance that we expect at the classical level under arbitrary change of reference
frame. Therefore, we further restrict the dependence of the wave-function $\sigma$ on the arguments so that it has simultaneously \emph{left and right gauge invariance}:
\begin{equation}\la{left-right-inv}
\sigma(\gamma_{\va L} q_{v} \gamma_{\va R}) = \sigma(q_v), \qquad \forall \gamma_{\va L},\gamma_{\va R}
\in \SU(2)
\end{equation}
There is an obvious consequence of this restriction: the field operator $\hsigma(h_v)$, now,
inherits gauge invariance on the left:
\begin{equation}\la{left-inv}
\hsigma(h_v) = \hsigma(\gamma h_{v})\,,~ \forall \gamma \in \SU(2).
\end{equation} 
The impact of this gauge invariance on the gluings is minimal. Indeed, one can explicitly check
that the convolutions of such wave-functions will still result in the identification of the bivectors
associated to the faces to be glued, by considering the expectation values of the scalar products
of the fluxes involved (which are indeed invariant under the rotations of the frames).

These operators $\hsigma,\hsigmad$ will allow us to construct state which incorporate
the vertex homogeneity idea in a very straightforward way, as they neatly separate the problem
of specifying the connectivity of the state from the implementation of vertex homogeneity. 
In other words, the states that we are going to build are based on the generic template:
\begin{equation}
\left(\int \prod_{v\in \Gamma}\extd h_v \prod_{i,j\in \Gamma} \delta(h_{v_i} h_{v_j}) 
%\prod_{v\in \Gamma} 
\hsigmad(h_{v}) \right)
\ket{0}\,,
\end{equation}
where the pattern of identifications implied by the product of Dirac deltas will encode
the structure of the graph that is required. 

Thus, for instance, a dipole state, encoding the simplest triangulation of $S^3$, will have the form
\begin{equation}\la{dipole}
\ket{\text{dipole}} \propto \int \extd h_v \extd h_w
\prod_{i=1}^{4} \delta(h_{(v,i)}h_{(w,i)}^{-1})
\hsigmad(h_{v})\hsigmad(h_{w})\ket{0}\,,
\end{equation}
while the triangulation generated taking five tetrahedra with the combinatorics of the boundary
of a $4-$simplex can be used for the state
\begin{align}
\ket{S^3}  
\propto& \int dh_{v_1} 
\extd h_{v_2} 
\extd h_{v_3} 
\extd h_{v_4} 
\extd h_{v_5} \nonumber \\
&\times
\delta(h_{(v_{1},1)}h_{(v_{5},{4})}^{-1})
\delta(h_{(v_{1},2)}h_{(v_{4},{3})}^{-1})
\delta(h_{(v_{1},3)}h_{(v_{3},{2})}^{-1})
\delta(h_{(v_{1},4)}h_{(v_{2},{1})}^{-1})
\delta(h_{(v_{2},2)}h_{(v_{5},{3})}^{-1}) \nonumber
\\&\times
\delta(h_{(v_{2},3)}h_{(v_{4},{2})}^{-1})
\delta(h_{(v_{2},4)}h_{(v_{3},{1})}^{-1})
\delta(h_{(v_{3},3)}h_{(v_{5},{2})}^{-1})
\delta(h_{(v_{3},4)}h_{(v_{4},{1})}^{-1})
\delta(h_{(v_{4},4)}h_{(v_{5},{1})}^{-1}) \nonumber\\
&\times
\hsigmad(h_{v_1})
\hsigmad(h_{v_2})
\hsigmad(h_{v_3})
\hsigmad(h_{v_4})
\hsigmad(h_{v_5})
\ket{0}\,.
\end{align}

As it is evident from these simple examples, the manipulation of complex states becomes
rapidly very difficult. However, we will now show how we can at least generate and control
some nontrivial states by introducing suitable refining operators.

\subsection{Simplicial moves}

Besides vertex homogeneity, the states that we seek will have to incorporate some sort
of information about the topology of the graphs, as well as some form of sum over geometries,
which is in fact the true novelty of the present proposal.

According to the kind of topological information that we are interested into, we might have to
consider GFT with additional data, in absence of which the reconstruction of the topological
properties encoded in a given graph might be ambiguous (this extra structure will be introduced in Section \ref{sec:melonic} by means of colored graphs). However, apart from this subtle
but important point, the construction follows exactly the same lines. 

We start from a certain seed state, built gluing $\sigma-$vertices with a combinatorial
pattern of our design (for instance a simple triangulation of a three sphere, as in the previous examples, or of a torus).
Then, we act on this state with simple refinement move operators, which have to be
\begin{itemize}
\item topology preserving;
\item respecting the vertex homogeneity.
\end{itemize}

The construction of the refinement move is rather constrained, given the requirements that
 need to be satisfied, and the peculiarities of the algebra of Bosonic ladder operators.
First, we implement the restriction on the preservation of topology. 
Natural candidates are operators realizing the basic Pachner moves ($1\rightarrow 4$,
$2\rightarrow 3$, $3\rightarrow 2$, $4\rightarrow 1$). In each of them, a certain number
of vertices will be destroyed and replaced, maintaining the topology, by a different number
of vertices.

We first notice that we are interested, in fact, into the \emph{growth} of a state out of a seed,
as we wish to start from states containing a minimal amount of vertices.
Second, due to the fact that the vertices that we are using are identical particle, it is impossible
to realize \emph{local} Pachner moves of the type $2\rightarrow 3$, $3\rightarrow 2$ and $4\rightarrow 1$, as it is impossible, without introducing explicitly some physical labelling of the
vertices (\eg through matter fields) to ensure that the vertices that are being destroyed are in fact
connected among themselves. 

On one hand, then, we are unable to explore ergodically the space of the possible triangulations
that can be reached with a sequence of Pachner moves starting from a given seed state. On the
other hand we are not necessarily interested in the implementation of this possibility, at this stage. In any case, this is just a matter related to the simplicity of the models that we are considering for the procedure to be applied, rather than a fundamental obstruction to the general idea. 

Therefore, we focus our attention onto the definition of a $1\rightarrow 4$ move, implemented
through the action of a certain composite operator which obeys a specific equation which we now
discuss. 
Being a $1\rightarrow 4$ move, the operator will involve an annihilator and four creation operators, convoluted with a kernel to be determined:
\begin{equation}
\move \equiv \int \prod_{i=1}^5 \extd g_{v_i} \MM(g_{v_1},g_{v_2},g_{v_3},g_{v_4};g_{v_5})
\hphid(g_{v_1})
\hphid(g_{v_2})
\hphid(g_{v_3})
\hphid(g_{v_4})
\hphi(g_{v_5}).
\label{move1to4}
\end{equation}
Notice that the kernel $\MM$ is completely symmetric in the first four sets of arguments only. The field operators have been chosen to be normal ordered as otherwise spurious $0\rightarrow 3$ terms, associated then to
disconnected components, would be
included.

The action of the operator onto a given state will generate a sum of terms, each of which will
correspond to a Wick contraction of the annihilator contained in $\move$ with one of
the creation operators used to build the state from the Fock vacuum, and the subsequent
creation of four new vertices with connectivity and data implied by $\move$.

In order to satisfy our requirements about the topology and the vertex homogeneity, then, we need the operator $\move$ to satisfy the equation
\ba\la{sim-move}
\left[ \move, \hsigmad(h_{1},h_2,h_3,h_4) \right] = \int dh_5\dots dh_{10}\,&&
\hsigmad(h_4,h_5,h_6,h_7) \hsigmad(h_7, h_3,h_8,h_9)\n\\
&&\hsigmad(h_9,h_6,h_2, h_{10})
\hsigmad(h_{10},h_8,h_5,h_{1})
\ea
where we recognize in the RHS the pattern of convolutions of four tetrahedra, with wave-function $\sigma$, glued to form a triangulation of $B^3$, which is exactly the same topology of the tetrahedron that has been removed. Graphically, the simplicial move action \eqref{sim-move} is
\be
\move:~~
\begin{array}{c}
\includegraphics[width=3.5cm]{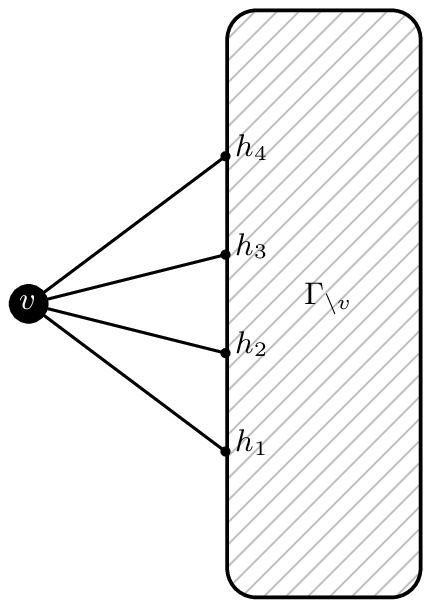}
\end{array}~~~~~\rightarrow~~~~~
\begin{array}{c}
\includegraphics[width=5cm]{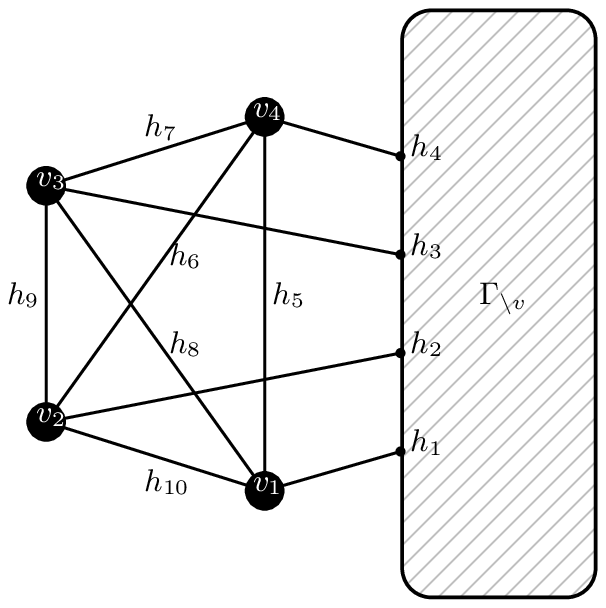}
\end{array}\,.
\ee

When using \eqref{move1to4}, this equation becomes a linear integral equation for the
kernel $\move$, for a fixed $\sigma$, which, we remember, will be fixed by the dynamics of
the system. 

Due to the non-abelian nature of $\SU(2)$ it is difficult to present a complete set of solutions,
with accessory conditions on $\sigma$. However, we claim that the procedure has a non-empty
set of solutions. 

To convince ourselves of this, we consider first the case in which the group is
abelian (\eg $\mathrm{U}(1)$). In this case, one can simply consider the equation for the kernel,
written in terms of representations, labeled by integers:

\begin{align}
&\MM_{
m_{1}^{1},m_{2}^{1},m_{3}^{1},m_{4}^{1},
m_{1}^{2},m_{2}^{2},m_{3}^{2},m_{4}^{2},
m_{1}^{3},m_{2}^{3},m_{3}^{3},m_{4}^{3},
m_{1}^{4},m_{2}^{4},m_{3}^{4},m_{4}^{4},
-m_{1}^{5},-m_{2}^{5},-m_{3}^{5},-m_{4}^{5}}
\sigma_{m_{1}^{5},m_{2}^{5},m_{3}^{5},m_{4}^{5}}\nonumber
%\label{LHSmain}
\\
&\times
\delta(m_{1}^{1}+m_{2}^{1}+m_{3}^{1}+m_{4}^{1})
\delta(m_{1}^{2}+m_{2}^{2}+m_{3}^{2}+m_{4}^{2})\nonumber
\\
&\times
\delta(m_{1}^{3}+m_{2}^{3}+m_{3}^{3}+m_{4}^{3})
\delta(m_{1}^{4}+m_{2}^{4}+m_{3}^{4}+m_{4}^{4})
\delta(m_{1}^{5}+m_{2}^{5}+m_{3}^{5}+m_{4}^{5})\nonumber
\\
=&\sigma_{m_{4}^{4},m_{2}^{4},m_{3}^{4},m_{4}^{4}}
\sigma_{m_{1}^{3},m_{3}^{3},m_{3}^{3},m_{4}^{3}}
\sigma_{m_{1}^{2},m_{2}^{2},m_{2}^{2},m_{4}^{2}}
\sigma_{m_{1}^{1},m_{2}^{1},m_{3}^{1},m_{1}^{1}}\nonumber
\\
&\times
\delta(m_{1}^{1}+m_{2}^{1}+m_{3}^{1}+m_{4}^{1}) 
\delta(m_{1}^{2}+m_{2}^{2}+m_{3}^{2}+m_{4}^{2})\nonumber
\\
&\times
\delta(m_{1}^{3}+m_{2}^{3}+m_{3}^{3}+m_{4}^{3})
\delta(m_{1}^{4}+m_{2}^{4}+m_{3}^{4}+m_{4}^{4})\nonumber
\\
&\times
\delta(m_4^{5}-m_1^{4})
\delta(m_3^{5}-m_2^{3})
\delta(m_2^{5}-m_3^{2})
\delta(m_1^{5}-m_4^{1})
\nonumber
\\
&\times
\delta(m_2^{4}+m_{3}^{1})
\delta(m_3^{4}+m_{2}^{2})
\delta(m_4^{4}+m_{1}^{3})
\delta(m_3^{3}+m_{2}^{1})
\delta(m_4^{3}+m_{1}^{2})
\delta(m_{4}^{2}+m_{1}^1)\,,
\end{align}
where no summation over repeated indices is implied.
Some explanations on the groups of Kronecker deltas are in order. 
 In the first line the spin flips are due
to the fact that there is a convolution between the kernel $\MM$ and $\sigma$. The fifth and the sixth lines  
come
from the gauge invariance of the four $\hphid$ that we will obtain in the commutator.
The seventh line follows from the relations among spins, implied by the dependence of the 
commutator on the group elements $h$ that are used for the gluing to the rest of the graph.
Finally,  
the last line comes from the gluings determined by the desired combinatorics
of the new piece to be attached to the state. 

Looking at the space of solutions of the equations implied by the Kronecker deltas, is it possible to show that the only delta in the LHS that is left unmatched, the one involving the indices $m_{i}^{5}$ and establishing the closure of the ``fluxes'' of the vertex labeled with $5$, is implied by
the other deltas on the RHS.
This implies that the desired condition is that $\sigma$ has to have nonzero (but otherwise arbitrary) components over all the representations (modulo gauge invariance). If this restriction on $\sigma$ is
imposed, then, we can simply divide the equation by
$\sigma_{m_1^{5}m_{2}^{5}m_{3}^{5},m_{4}^5}$ and get our solution for the kernel.

Notice that here $m$ plays the role of what would be $B \in \mathfrak{su}(2)$ if we were considering the nonabelian case. In fact, the nonabelian equation can be obtained by the
replacement $m_{i}^{v} \rightarrow B_{i}^{v}$ and with $\cdot \rightarrow \star$. Unfortunately,
in the case of the $\star-$product, there is no obvious counterpart of the division.

If we keep working with $\SU(2)$, though, we can show that there is at least one special
class of wave-functions $\sigma$ that allows us to determine explicitly $\move$. 
If we impose the rather demanding (in the sense of restriction on the family of states allowed by our construction) condition:
\begin{equation}
\left[ \hsigma(h_v), \hsigmad(h_w) \right] = \Delta_{L}(h_v, h_w) \equiv \int_{SU(2)} d\gamma 
	\prod_{i=1}^4 \delta(\gamma h_{(v,i)} h_{(w,i)}^{-1}), \label{restriction}
\end{equation}
where the \emph{left} invariant delta function appears due to the symmetry properties of the operators
(and again we used the notation   $h_v = (h_{(v,1)},h_{(v,2)},h_{(v,3)},h_{(v,4)})$),
one can see very easily that
\begin{equation}
\move = \int dh_{1}\dots dh_{10}\, 
\hsigmad(h_{4},h_{5},h_{6},h_{7})
\hsigmad(h_{7},h_{3},h_{8},h_{9})
\hsigmad(h_{9},h_{6},h_{2},h_{10})
\hsigmad(h_{10},h_{8},h_{5},h_{1})
\hsigma(h_{1},h_{2},h_{3},h_{4})
\end{equation}
is a solution for our equation. We refer to the appendix \ref{app:calculationsrefinement} for the detailed
discussion of the restriction on the wave-function in terms of its components.

We should stress that this restriction on $\hsigma$ should not be seen as a statement about
the uniqueness of the solution to our equation for $\move$. Indeed, this is just
a sufficient condition to have a refinement move operator, not a necessary one. 
As the abelian case
has shown, it is likely that there are many more pairs of $\hsigma,\move$ which simply
we cannot present explicitly due to technical difficulties intrinsic in the nonabelian nature
of $\SU(2)$ (or $\su(2)$). Moreover, the conditions we find for satisfying \eqref{restriction} are quite general, i.e. they lead to an ample family of solutions.

It is immediate to show that, for instance, the states presented in the previous subsection
are related by a refinement move
\begin{equation}
\move\ket{\text{dipole}} \propto \ket{S^3} \,,
\end{equation}
as expected.

%%%%%%%%%%%%%%%%%%%%%%%%%%%%%%%%%%%%%%%%%%%%%%%%%%%%%%%%%%%%%%%%%%%%%%%%%%%%%%%%%%%%%%

%%%%%%%%%%%%%%%%%%%%%%%%%%%%%%%%%%%%%%%%%%%%%%%%%%%%%%%%%%%%%%%%%%%%%%%%%%%%%%%%%%%%%%

\section{Graph-encoded topologies}
\label{sec:GraphTop}

\subsection{Generalities of these graphical structures}

\begin{figure}[htb]
   \centering
   \includegraphics[scale=1]{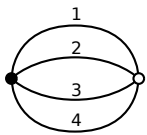}
   \caption{\label{fig:supermelon} A very simple $4$--colored graph.}
 \end{figure}

 A \emph{closed $4$--colored graph} $\bgraph$ is a bipartite, $4$--regular, edge--colored graph.  
 In this context, \emph{edge--coloring} has a precise meaning: each edge of $\bgraph$ is marked with a color drawn from the set $\{1,2,3,4\}$, such that the four edges incident at any given vertex are marked with distinct colors. 
 A simple example is drawn in Figure \ref{fig:supermelon}.

 A \emph{$k$--bubble of species $\{i_1,\dots,i_k\}$} is a maximal connected, $k$--colored subgraph $\cset^{i_1\dots i_k}\subseteq\bgraph$, with color set $\{i_1,\dots,i_k\}\subseteq \{1,2,3,4\}$. 

Thus, in the context of 4--colored graphs, there are 0--, 1--, 2--, 3--bubbles as well as a single 4--bubble (the graph itself). These bubbles play respective topological role as vertices, edges, faces, three--dimensional cells and the manifold itself.  Moreover, the bubbles form a nested arrangement of structures within $\bgraph$: $(k-1)$--bubbles are nested within $k$--bubbles that are in turn nested within $(k+1)$--bubbles and so on.  

 \begin{figure}[htb]
   \centering
   \includegraphics[scale =0.3]{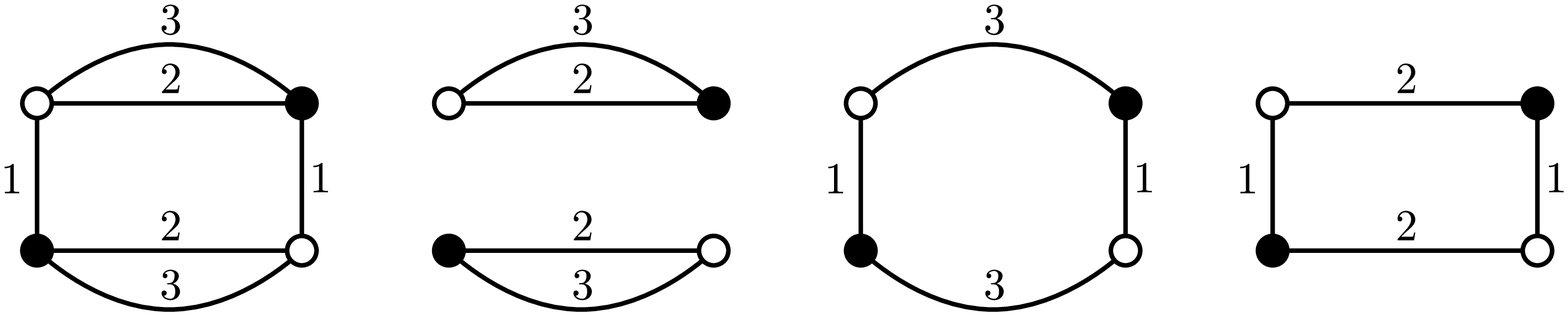}
   \caption{\label{fig:BubbleEx} A $3$--colored graph and its 2--bubbles.}
 \end{figure}

This presents an opportunity to comment on the manner in which $4$--colored graphs encode $3$--dimensional abstract simplicial pseudo--manifolds. Without dwelling on the details, one constructs a $3d$ abstract simplicial complex from the $4$--colored graph by:
\begin{itemize}
  \item[--] associating a $(3-k)$--simplex to each $k$--bubble in $\bgraph$;
  \item[--] in such a manner that the nested structure of the bubbles translates to a dual nested structure for the simplicial cells.  
\end{itemize}
One can show that such structures satisfy the triple of properties (pure, non--branching, strongly--connected) required to constitute a pseudomanifold.
In this work, interest concentrates on manifold structures, specifically $3$--spheres, $3$--balls and $3$--shells.

\vspace{1.0cm}

\begin{figure}[htb]
\centering
\includegraphics[scale =1]{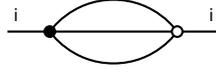}
\caption{\label{fig:onedipole}A melon of species \{i\}.}
\end{figure}

\emph{Melons} are central to a set of combinatorial moves on the graphs and will be used in Section \ref{sec:melonic} below. A melon is a graph of the form illustrated in Figure \ref{fig:onedipole}. 

\begin{figure}[htb]
\centering
\includegraphics[scale = 1]{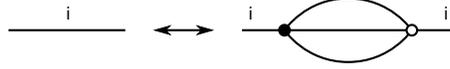}
\caption{\label{fig:dipolemove} Melon creation and annihilation.}
  \end{figure}

\emph{Melonic moves} come in two forms, creation and annihilation:
  \vspace{-0.2cm}
  \begin{itemize}
    \item[--]
      The \emph{creation} of a melon of species $\{i\}$, inserts such a melon along an edge of color $\{i\}$ in $\bgraph$.
\item[--]
  The \emph{annihilation} of a melon of species $\{i\}$ removes the melon and reroutes the edge of color $\{i\}$. It corresponds to a right--to--left passage in Figure \ref{fig:dipolemove}.
 
\end{itemize}

 While such moves are defined combinatorially, they are topology preserving in that two graphs differing by such a subgraph encode pseudomanifolds that are homeomorphic.

\section{Melonic moves}\la{sec:melonic}

In the previous section we defined graphical structures, (signed) 4-colored graphs, that encoded $3d$ (simplicial) topologies, as well as a set of topology-preserving (dipole) moves. Fortunately, these are well-adapted to the operatorial GFT approach and allow the construction of different types of refinement operators for greater control over the corresponding topological structures. 

Therefore, the storage of the topological properties of the dual triangulation requires the inclusion of such colouring in the GFT field operators defined above. One can also interpret each color label as defining a different GFT field, as we shall see. These additional data 
will allow a complete reconstruction of the dual simplicial complex determined by a GFT state.
This is particularly important for the problem we want to solve because the state has to correspond
to a triangulation with a specific topology and with a certain number of disconnected boundaries.

\subsection{Isolating the quanta} 

The building blocks of 4-colored graphs are 4-valent vertices carrying an additional discrete label, which has only two values: $B,W$. This will be represented in our
graphs by drawing vertices as black or white circles. Correspondingly, the outgoing strands of
each vertex will be numbered $1,2,3,4$ clockwise for black vertices and anticlockwise for white
ones. This number will correspond to the assignment of a specific colour to the corresponding
strand. Graphically, we can depict the action of the color fields as
\be\la{TWB}
\hphid_{\va W}(g_{1},g_{2},g_{3},g_4)  \ket{0}=
\begin{array}{c}
\includegraphics[width=2.5cm]{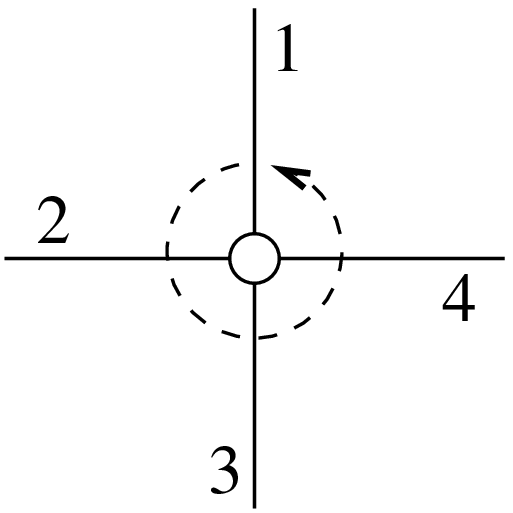}
\end{array}\,,~~~~~~
\hphid_{\va B}(g_{1},g_{2},g_{3},g_4) \ket{0}=
\begin{array}{c}
\includegraphics[width=2.5cm]{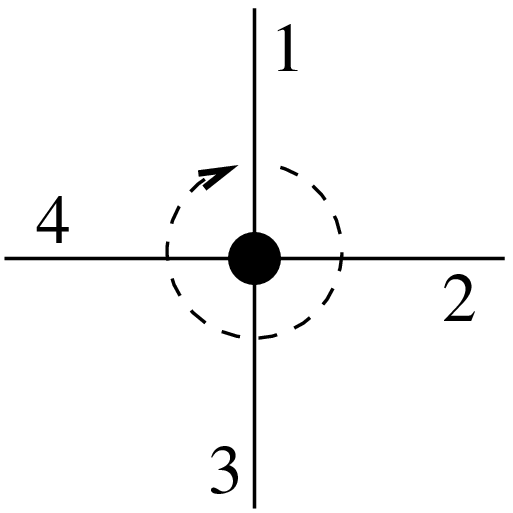}
\end{array}\,.
\ee

With these conventions, graphs will be built by gluing vertices according to two rules: 1) vertices
with the same colour cannot be connected by an edge; 2) only strands with the same colour can be
joined to form an edge.

The introduction of coloured graphs implies a (minimal) modification of the simple GFT models
discussed previously in this paper. The colouring of the vertices implies that the field operators
are now doubles, with $\hphi_{\va {B,W}},\hphid_{\va {B,W}}$ being now a family of ladder operators
for black and white vertices.
The colouring of the strands implies that the group variables
representing the argument of the field operators carry implicitly an additional (discrete) label,
the colour, which is essentially the number of the slot of the field operator in which they appear.
It is tempting to interpret these new data geometrically. In particular, the additional $B,W$ label
might be ultimately connected to the extension of GFT from the special case of $\SU(2)$ (that we consider in this paper) to the group $Pin(3)$, since the new labels correspond to opposite orientations that can be assigned to the same simplicial structures. However, the elaboration of a more precise statement is left for 
future work.

The commutator of the colored field operators reads
\begin{equation}\la{CCRright}
  [\hphi_{t(v)}(h_v),\hphid_{t(w)}(h_w)] = \delta_{t(v), t(w)} \int_{\SU(2)} d\gamma \prod_{i=1}^4 \delta( h_{(v,i)} \gamma h_{(w,i)}^{-1})\,,
\end{equation}
with $t(v)\in\{B,W\}$. It is with respect to these operators that we define the Fock vacuum $\fv$, such that
\begin{equation}
  \hphi_{\va B}(g_v)\fv = \hphi_{\va W}(g_v)\fv = 0\,.
\end{equation}

The new color $B,W$ will  be inherited by the redefined GFT field operator \eqref{field} encoding the vertex wave-function, so that our basic building blocks are now represented by
\be\la{c-field}
\hsigma_{t(v)}(h_v) = \int \extd g_v\; \sigma(h_vg_v) \,\hphi_{t(v)}(g_v)\,, \quad\quad \hsigmad_{t(v)}(h_v) = \int_{\SU(2)} \extd g_v\; \overline{\sigma}(h_vg_v) \,\hphid_{t(v)}(g_v)\,,
\ee
 satisfying the commutation relations
\be
\left[\hsigma_{t(v)}(h_v), \hsigmad_{t(w)}(h_w) \right] = \delta_{t(v), t(w)}\Delta_{L}(h_v, h_w) \equiv  \delta_{t(v), t(w)} \int_{\SU(2)} d\gamma \prod_{i=1}^4 \delta(\gamma h_{(v,i)} h_{(w,i)}^{-1})\,, \label{c-restriction}
\ee
encoding the left gauge invariance of the vertex wave-function \eqref{left-inv}.

\subsection{Closed graph-labelled operators}

At last, we are in a position to construct GFT states with support on closed graphs. Generically, these take the form:
\begin{equation}
  \ket{\psi_{\bgraph}} = \int [dh_v]_\vset \;  \kernel_\bgraph\big(\{h_{(v,i)}\}_\eset\big)\prod_{v\in\vset}\hsigmad_{t(v)}(h_{v})\fv\,,
\end{equation}
where $\bgraph$ is the graph that provided support for the state, $\vset$ and $\eset$ are its vertex and edge sets,  $\kernel_{\bgraph}$ is a graph-adapted group kernel. By graph-adapted group kernel, we mean that $\kernel_\bgraph$ associates group arguments pairwise -- consider an edge $e$ of color $i$ in $\bgraph$ and say that its endpoints are the vertices $v$ and $v'$, then the two group arguments $h_{(v,i)}$ and $h_{(v',i)}$ are paired in the kernel $\kernel_\bgraph$.  In this instance, we shall pick a very simple kernel:
\begin{equation}
  \kernel_{\bgraph}\big(\{h_{(v,i)}\}_{\eset}\big) = \prod_{i = 1}^4 \prod_{e = (vv')\in \eset_i} \delta\big(h_{(v, i)}, h_{(v',i)}\big)\,,
\end{equation}
where $\eset_i$ is the set of edges of $\bgraph$ of color $i$.
Of course, we could pick something other than a $\delta$-function.  
Comparing our choice here with the generic wave-function laid out in Equation \eqref{eq:GenWave}, one sees that:
\ba
  \label{eq:Compare}
  \psi_{\bgraph}\big(\{g_v\}_{\vset}\big) &=& \int [\extd h_{v}]_{\vset}\; \kernel_{\bgraph}\big(\{h_{(v,i)}\}_{\eset}\big) \prod_{v\in\vset}
  \int \extd k_v\; \overline{\sigma}_{t(v)}(k_vh_vg_v)\n\\
  &=&\int [\extd h_{v}]_{\vset}\; \kernel_{\bgraph}\big(\{h_{(v,i)}\}_{\eset}\big) \prod_{v\in\vset}
  \overline{\sigma}_{t(v)}(h_vg_v)\,.
\ea
where $k_v\in\SU(2)$ parameterises the projection to the gauge-invariant subdomain\footnote{To be clear: $\overline{\sigma}_{t(v)}(k_vh_vg_v) =  \overline{\sigma}_{t(v)}\big(k_v h_{(v,1)}g_{(v,1)},k_v h_{(v,2)}g_{(v,2)},k_v h_{(v,3)}g_{(v,3)},k_v h_{(v,4)}g_{(v,4)}\big)$.} and we have expressed explicitly the left gauge invariance of the vertex wave-function $\sigma_{t(v)}(h_vg_v)$.

It is also handy to define operatorial quantities with support on open graphs. An open 4-colored graph has interior and exterior substructure $\bgraph = \bint\cup\bext$.  The state assigned to such a graph has the form:  
\begin{equation}
  \ket{\psi_{\bgraph}\big(\{h_{(v,i)}\}_{\eext}\big)} = \int [\extd h_{(v,i)}]_\eint \;  \kernel_{\bint}\big(\{h_{(v,i)}\}_{\eint}\big)\prod_{v\in\vint}\hsigmad_{t(v)}(h_{v})\fv\,,
\end{equation}
where:
\begin{equation}
\kernel_\bint\big(\{h_{(v,i)}\}_{\eint}\big) =  \prod_{i = 1}^4 \prod_{e = (vv')\in \einti} \delta\big(h_{(v,i)}, h_{(v',i)}\big)\,.
\end{equation}
Thus, the state assigned to such a graph will be parameterised also by the group elements associated to the exterior edges.

A yet more subtle issue is the realization of graph manipulation in this operatorial context. This is imperative for the refinement moves that we wish to utilize.  
There are two basic operations on a state, the action of $\hsigmad_t$ and $\hsigma_t$. The former effectively acts by multiplication, creating a vertex of the appropriate color. Meanwhile, the latter effectively acts by derivation, annihilating each one of the existing vertices in turn to yield of series of states:
\begin{equation}
  \hsigma_{t(u)}(h_{u})\ket{\psi_{\bgraph}} = \sum_{\substack{v\in\vset\\[0.05cm] t(v) = t(u)}} \ket{\psi_{\bgraph\bs v}(h_u)} \,,
\end{equation}
where $\bgraph\bs v$ is the open graph obtained from $\bgraph$ by cutting out the vertex $v$. 
This is the basic mechanism by which one can manipulate GFT states. 

We shall assume that the information about the GFT dynamics is captured by states with support on closed graphs.\footnote{This assumption is well motivated.  While the current level of knowledge about the GFT dynamics does not permit one to state definitively that closed graphs capture exclusively the non-trivial GFT information, one would expect that states based on open 4-colored graphs are generically annihilated by the GFT dynamics, except when their exterior edge group arguments accidentally coincide to effectively close the graph. This stems from the fact that open graphs are not gauge-invariant.}  As a result, we consider only those manipulating operators that ultimately maps from closed graphs to closed graphs.    This entails that the class of relevant, normal ordered operators, can be catalogued\footnote{This cataloguing is unique up to graph automorphism in the following sense. If the graph-adapted group kernel is invariant under the action of a graph automorphism, then configurations related by that automorphism give rise to the same operator.}   by closed 4-colored graphs, a choice of graph-adapted group kernel and a choice for each vertex whether to use a creation or annihilation operator. In symbols, a generic operator on Fock space has the form:
\begin{equation}
  \op_{\opgraph,\opkernel,\choice} = \int [\extd h_v]_\vset \;\opkernel_{\opgraph} \big(\{h_{(v,i)}\}_{\eset}\big) \prod_{\substack{v\in\vset\\[0.05cm] c(v) = 1}}\hsigmad_{t(v)}(h_v)\prod_{\substack{v\in\vset\\[0.05cm] c(v) = -1}} \hsigma_{t(v)}(h_v)\;, 
\end{equation}
where $\opgraph$ is a closed 4-colored graph, $\opkernel_\bgraph$ is the graph-adapted group kernel and $\choice = \{c(v) : v\in\vset\}$, where $c(v)=1$ associates a creation operator to $v$ and the $c(v)=-1$ associates an annihilation operator to $v$. In this instance, we shall again choose a very simple kernel for these operators:
\begin{equation}
  \opkernel_{\opgraph}\big(\{h_{(v,i)}\}_{\eset}\big) = \prod_{i = 1}^4 \prod_{e = (vv')\in \eset_i} \delta\big(h_{(v, i)}, h_{(v',i)}\big)\,.
\end{equation}
The action of such an operator on a state yields:
\begin{equation}
  \op_{\opgraph,\opkernel,\choice}\ket{\psi_\bgraph} = \sum_{\gamma} \ket{ \psi_{(\opgraph\bs\vset^-)\#_{\gamma}(\bgraph\bs\gamma(\vset^-))} } 
\end{equation}
where $\vset^- = \{v\in\vset_\opgraph:c(v) = -1\}$ and $\gamma:\vset^-\rightarrow \vset_\bgraph$ such that $\gamma(t(v)) = t(v)$. In other words, $\gamma$ maps the vertices in $\vset^-\subset\vset_\opgraph$ to a subset of vertices $\gamma(\vset^-)\in\vset_\bgraph$ preserving the species (\{B,W\}) of the vertex. The term $\#_\gamma$ denotes that the open edges in $\opgraph$ created by the excision of $v\in\vset^-$ are glued to the open edges in $\bgraph$ created by the excision of $\gamma(v)\in\gamma(\vset^-)$.  

For example, consider the number operator
\ba\la{ numop}
  \numop_t &=& \op_{\opgraph_{\va v_1}, I,\{1,-1\}} = \int \extd h_v\,\extd h_{v'} \left[\prod_{i = 1}^4\delta(h_{(v,i)},h_{(v',i)})\right]  \hsigmad_{t(v)}(h_v) \, \hsigma_{t(v)}(h_{v'})\n\\
  &=& \int\extd h_v \; \hsigmad_{t(v)}(h_v)\,\hsigma_{t(v)}(h_v)\,,
\ea
where $\opgraph_{\va v_1}$ is the supermelon graph in FIG. \ref{fig:supermelon}.

Using these tools one can easily build graphs that are dual to the desired topology. 
The general strategy is the same as in the case of simplicial moves. We start
from a state, playing the role of a seed;
then, we will act with an operator that realize a very simple move that changes the
triangulation leaving fixed the topology. In this way we will be able to grow in an arbitrary
way a large subclass of all the possible triangulations compatible with the given topology. 
 Let us now provide some examples of this construction.

%%%%%%%%%%%%%%%%%%%%%%%%%%%%%%%%%%%%%%%%%%%%%%%%%%%%%%%%%%%%%%%%%%%%%%%%%%%%%%%%%%%%%%

%%%%%%%%%%%%%%%%%%%%%%%%%%%%%%%%%%%%%%%%%%%%%%%%%%%%%%%%%%%%%%%%%%%%%%%%%%%%%%%%%%%%%%

\section{3-Sphere}
\la{sec:sphere}
The first example is also the simplest closed topology one can consider, but of direct use in (quantum) cosmology applications.

\subsection{Seed state}

\begin{figure}[htb]
\includegraphics[height=4cm]{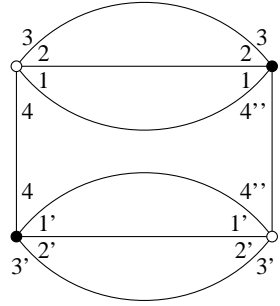}
\caption{Two nested melons representing the dual graph to a possible triangulation of a 3-sphere.}
\label{bimelon}
\end{figure}

A state associated to a simple triangulation of the 3-sphere can be built from the (dual) graph represented in FIG. \ref{bimelon}
as:
\begin{equation}\la{seed-sphere}
\ket{S^3} =\int (dg)^{8}
\hsigmad_{\va W}(g_{1},g_{2},g_{3},g_{4})
\hsigmad_{\va B}(g_{1'},g_{2'},g_{3'},g_{4})
\hsigmad_{\va W}(g_{1'},g_{2'},g_{3'},g_{4''})
\hsigmad_{\va B}(g_{1},g_{2},g_{3},g_{4''})
\fv\,,
\end{equation}
where from now on $(dg)^n$ indicates the total Haar measure associated to the $n$ group elements appearing as arguments of the field operators in the integrand.

\subsection{Refinement}

The state \eqref{seed-sphere} can be used as a seed to generate more and more refined triangulations of the 3-sphere  by repeated action of a single refinement operator.  The general idea is to replace a vertex of the original triangulation with a group of vertices such that the topology is kept fixed, in a way that closely
follows the Pachner moves.

The simplest melon, built with two nodes, will not do the job, as it would result into the replacement of one vertex with another copy of the same vertex. Therefore, the simplest
move involves at least two nested melons, as in FIG. \ref{bimelon}, and we need to consider again the operator entering the definition of our seed state \eqref{seed-sphere}.
However, we now replace one of the creation operators in  \eqref{seed-sphere} with an annihilation operator for the field with opposite color. For instance, a second quantized operator which act on the states and change
them with the desired moves can be written as
\ba\la{move-circ}
\move_{\va W}&\equiv  &\move (W\rightarrow WBW) \nonumber\\  
& =&  \int (dg)^8
 \hsigmad_{\va W}(g_{1},g_{2},g_{3},g_{4}) 
 \hsigmad_{\va B}(g_{1'},g_{2'},g_{3'},g_{4})
\hsigmad_{\va W}(g_{1'},g_{2'},g_{3'},g_{4''})
\hsigma_{\va W}(g_{1},g_{2},g_{3},g_{4''})\,;
\ea
such operator replaces one white vertex with three vertices, two white and one black, in such a way that the legs that were glued to the original white
vertex are now glued to the two new white vertices. 
 Using the algebra of ladder operators, this can be seen from the commutator:
\begin{align}
[\move_{\va W},\hsigmad_{\va W}(h_{1},h_2,h_3,h_4)] = 2\int (dg)^4 
\hsigmad_{\va W}(h_{1},h_2,h_3,g_4) \hsigmad_{\va B}(g_{1'},g_{2'},g_{3'},g_4) \hsigmad_{\va W}(g_{1'},g_{2'},g_{3'},h_4).
\end{align}
Graphically, we have
\be\la{move-circ}
\move_{\va W}:~~
\begin{array}{c}
\includegraphics[width=1.7cm]{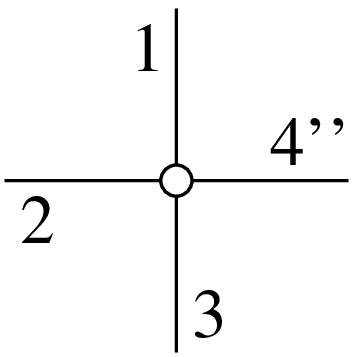}
\end{array}~~~~~\rightarrow~~~~~
\begin{array}{c}
\includegraphics[width=5.7cm]{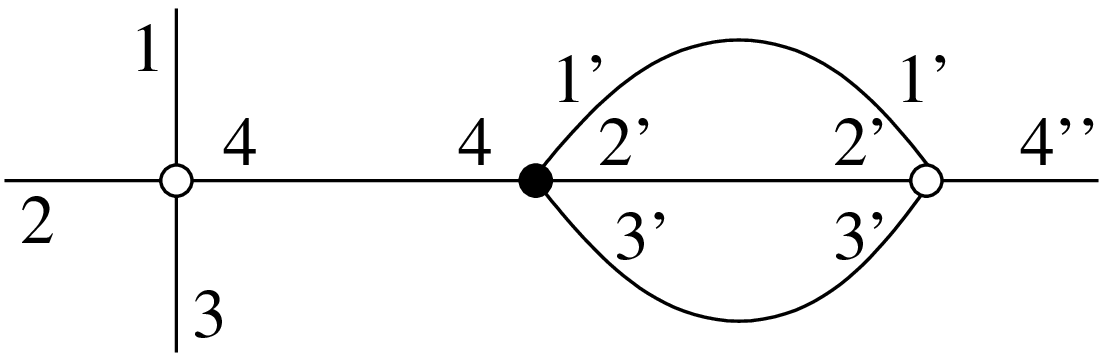}
\end{array}\,.
\ee
Notice that, as the states are constructed with strings of creation operators acting on the Fock vacuum, the move is inactive on black vertices,
as $[\move_{\va W},\hsigmad_{\va B}] = 0$. 

Due to the symmetry of the melonic graph at hand, we can build only two moves, one acting on the white vertices and another on the black ones (obtained with an analogous construction where instead we replace one of the $\hsigmad_{\va W}$ in \eqref{seed-sphere} with $\hsigma_{\va B}$).
More complicated graphs might result into a larger set of them.
However, it is natural to expect that at least some, if not all, of them will be simply reducible to the action of compositions of these basic moves.

%%%%%%%%%%%%%%%%%%%%%%%%%%%%%%%%%%%%%%%%%
%%%%%%%%%%%%%%%%%%%%%%%%%%%%%%%%%%%%%%%%%

\section{Spherical Shell}
\la{sec:shells}
The next example is more involved (thus, also more interesting) due to the presence of boundaries, which require special care. It will be also the basic topological building block for constructing more intricate topologies as well as states associated to less symmetric quantum geometries. 

Indeed, the construction of the previous section can be immediately generalised to cases in which the symmetry group is not acting transitively on the manifold, but it is
instead defining a foliation of the manifold into homogeneous orbits. This is the
case, for instance, of spherically symmetric geometries, as we will discuss again briefly in the concluding section.

The leaves of such foliations will correspond to triangulations of spherical shells (thus fixing the topology to be $S^2\times [0,1]$
), with geometric data
assigned in such a way to have an isometry group which is $\SO(3)$.

\subsection{General properties}
As a first step, we concentrate on a single 3-shell here. 
The shell will be generated by acting with a suitable refinement operator upon an appropriately designed seed state, which will be obtained from the simplest triangulation of a shell with two boundaries (`internal' and `external'). As for the case of the 3-sphere then, the action of an arbitrary number of refinement moves will generate (dynamically) larger and larger triangulations of the shell.

While the assignment of the geometric data to the vertices is done through the
wave-functions attached to the vertices of the graph (through the
construction of suitable field operators encoding them directly as we have
seen in the previous section), the topology of the state is controlled by
the topology of the graph. The problem we need to address is to generate arbitrary triangulations of the spherical shell.

\begin{figure}
\includegraphics[height=4cm]{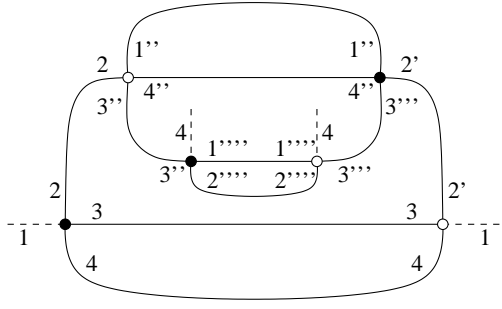}
\caption{Three nested melons with open edges can be shown to be dual to
a shell with two simple spherical boundaries, triangulated with two triangles each.}
\label{trimelon}
\end{figure}

The general methods discussed so far are able to completely control the topology of the triangulation dual to a graph. They are still not enough, for our purposes. 
Indeed, as one can easily see from Fig. \ref{trimelon} corresponding, in four dimensions, to the simplest triangulation of a spherical shell with two boundaries, we can
associate to each boundary the colour of the links dual to it. However, because of the
particular structure of the graphs, the colours of the boundary links will appear also
in the bulk of the shell. While there is nothing wrong with this, in general, when computing the action of
certain geometric operators (see Section \ref{sec:geometry}) we would like to be able to distinguish the contributions of the links connecting different shells from links within a single shell. This is due to the fact that
the kind of operators that we are interested in will be one-body operators,
acting on each graph on each single vertex, and hence unable to detect the global properties
of the graph the vertex belongs to. This requires more data than the colours alone.

On top of the standard colouring, we then need to associate an additional label  to each vertex, which will tell us whether the vertex is adjacent to a boundary or not. We take this new label $s$ to take values $s=+,0,-$. 

One option is to attach this new label to the field operators themselves, defining the operators $\hphid_{\va B+}, \hphid_{\va W+} $ which create respectively black  and white vertices with an open link  belonging to one of the external boundaries of the shell.  
Analogously, the operators $\hphid_{\va B-}, \hphid_{\va W-}$ create black  and white vertices  with an open link belonging to the other boundary.  
Finally,  the operators $\hphid_{\va B0}, \hphid_{\va W0}$ create internal vertices with no open links. In the graphic representation \eqref{TWB}, the associated values of the new label $s$ are to be added at each vertex.

This procedure, while convenient, is however rather unsatisfactory from the point of view of the microscopic theory we are starting from. Indeed, these labels are significant only for particular states,
as they are used to identify special \emph{loci} within a special class of graphs.
Therefore instead of introducing this additional label on the fundamental operators (which are still carrying a B/W label), we move it on the wave-functions $\sigma$.

While this changes little at the level of the computations, from a conceptual point of view this is much more satisfactory as the  $\hsigma$'s are the operators that are adapted to the
special class of states. Therefore, there is no problem for them to carry additional structure
that refers precisely to those states.
 Consequently, the field operators to be used in the construction of our states can be written as
\be\la{cfield}
\hsigma_{ts}(g_{v}) = \int \extd h_v \sigma_{s}(g_v h_v)\hphi_{t}(h_v)=\int \prod_{i=1}^4 dh_i \sigma_{s}(g_1 h_{1},g_{2}h_{2},g_{3}h_{3},g_{4}h_{4}) 
\hphi_{t}(h_{1},h_2,h_3,h_4)\,,
\ee
with $t=B,W, s=+,0,-$.  This definition, now, requires that the wave-functions $\sigma_{s}$ are
orthogonal under convolution even if they have a different label $s$\footnote{This additional label, at this stage, is introduced by hand. It has to be understood as a shorthand notation for the result of an operation
of identification of the desired region of the graph in terms of global properties. In particular,
it may stand for the procedure that would consist, in LQG, in the identification of a surface with its implicit equation and then in the selection of those links of the graph intersecting it.}.

\subsection{Seed state}

We are now ready to construct the seed state for a shell  corresponding to the graph depicted in FIG. \ref{trimelon}. 
By arbitrarily choosing the internal boundary as corresponding to the open links 1 and the external boundary to the open links 4, we have
\ba\la{tau}
\ket{\tau} &=&
\int (dg)^{10} 
\hsigmad_{\va B-}(e,g_{2},g_{3},g_4)
\hsigmad_{\va  W-}(e,g_{2'},g_{3},g_4)
\hsigmad_{\va  B0}(g_{1''},g_{2'},g_{3'''},g_{4''})\n\\
&&~~~~~~~~~~~\hsigmad_{\va  W0}(g_{1''},g_{2},g_{3''},g_{4''})
\hsigmad_{\va B+}(g_{1''''},g_{2''''},g_{3''},e)
\hsigmad_{\va  W+}(g_{1''''},g_{2''''},g_{3'''},e)
\fv \, .
\ea

This state corresponds to a single shell, with two boundaries which, eventually, can be glued to the boundaries of other shells. A tedious calculation shows that indeed the state $\tau$ has a wave-function obtained with the desired pattern of convolutions.

\subsection{Refinement}

The state \eqref{tau}  can be used to define states on more general triangulations. In analogy to the 3-sphere case, this can be achieved by defining a class of moves that can be used to manipulate the states, translating a set
of corresponding manipulations of their associated graphs.
Again, in replacing a vertex of the original triangulation with a set of new vertices the topology has to be left unchanged.

There are many possible choices, of course. Here we consider just a minimal set of moves that are at our disposal and that can act on the desired states.
The proof of the general properties of these moves (for instance, whether they act transitively on the space of graphs with the same topology, \ie if they
are able to connect any pair of triangulations of the same shell) is left for future work, as it is not very relevant for the case at hand.

The structure of the graphs used so far implies that there will be (at least) three classes of refinement moves: 
\begin{enumerate}
\item refinement of the bulk (vertices with label $0$);
\item refinement of the boundary \lq $+$\rq;
\item refinement of the boundary \lq  $-$\rq.
\end{enumerate}
While we are mainly interested in the refinement of the boundaries (corresponding to two dimensional spheres triangulated with more and more triangles),
we consider also the bulk vertices, for completeness.

The construction of the refinement move, for all of these cases, follow the same logic. The idea is to attach to the initial graph $\Gamma_0$ another portion obtained
by a different graph $\Gamma_{add}$. This will be done by deleting from both graphs a vertex, $\Gamma_{0 \setminus v}$ and $\Gamma_{add \setminus v}$, and by gluing the resulting graphs along the new open edges, $\Gamma_1 = \Gamma_{0 \setminus v}\#_{v,v'}\Gamma_{add \setminus v'} $, 
so that the colouring rules are respected and the topology is left unchanged.

\subsubsection{Bulk vertices}

To refine the triangulation in the bulk, it suffices to consider the move that removes a bulk vertex and replaces it with the graph obtained removing one vertex from
a graph dual to the triangulation of the three sphere.

In our case, then, we can use melonic graphs, associated to vertices of colour zero only. In this way, the topology of the graphs is left unchanged. The vertices with colours $\pm$ would necessary introduce new boundary components, possibly disconnected from the previous ones. For this reason we shall not consider mixed configurations.

The construction of the refinement move for the bulk vertices follows exactly the one already described in the case of the 3-sphere. 
Hence, we give immediately the expression for the operator performing the refinement move replacing one white vertex with two white and one black ones, namely
\ba\la{W0}
\move_{\va W0}&\equiv  &\move_{0} (W\rightarrow WBW)  \nonumber\\  
&=&  \int (dg)^8
 \hsigmad_{\va W0}(g_1,g_2,g_3,g_4) \hsigmad_{\va B0}(g_{1'},g_{2'},g_{3'},g_4) \hsigmad_{\va W0}(g_{1'},g_{2'},g_{3'},g_{4''})
\hsigma_{\va W0}(g_1,g_2,g_3,g_{4''})\n\\
\ea
Again, the refinement move follows from the commutator:
\ba
&&[\move_{\va W0},\hsigmad_{\va W0}(h_{1},h_2,h_3,h_4)]\n\\
&&= 2\int (dg)^4 
\hsigmad_{\va W0}(h_{1},h_2,h_3,g_4) \hsigmad_{\va B0}(g_{1'},g_{2'},g_{3'},g_4) \hsigmad_{\va W0}(g_{1'},g_{2'},g_{3'},h_4)\,,
\ea
which graphically can be represented as 
\be
\move_{\va W0}:~~
\begin{array}{c}
\includegraphics[width=1.7cm]{M0l.eps}
\end{array}~~~~~\rightarrow~~~~~
\begin{array}{c}
\includegraphics[width=5.7cm]{M0r.eps}
\end{array}\,.
\ee

The analogous expression for  $\move_{\va B0}\equiv  \move_{0} (B\rightarrow BWB)$ can be immediately derived.

\subsubsection{Boundary vertices}

The construction of the refinement moves for the boundary vertices follows the same path. The move will be constructed out of an operator associated to a triangulation
of a ball. The simplest melonic graph (with open edges of colour $1$) dual to a ball can be easily seen to be:
\begin{align}
D_{\pm} &= \int dg_{2}
dg_{3}
dg_{4}
dg_{4'}
dg_{2''}
dg_{3''}
 \nonumber \\
&
\hsigmad_{W\pm}(e,g_2,g_3,g_4)
\hsigmad_{B\pm}(e,g_2,g_3,g_{4'})
\hsigmad_{W\pm}(e,g_{2''},g_{3''},g_{4'})
\hsigmad_{B\pm}(e,g_{2''},g_{3''},g_4)\,,
\end{align}
which we can graphically depict as 
\be\label{boundary-ball}
\begin{array}{c}
\includegraphics[width=6.5cm]{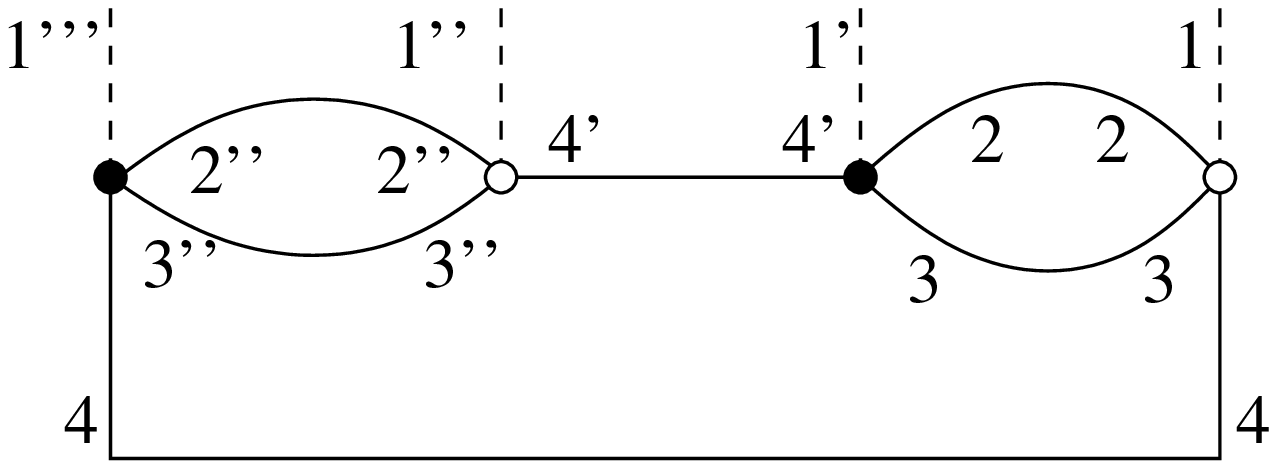}\,.
\end{array}
\ee
A similar operator can be built for any other choice of colour for the boundary edges.

It can be seen that the moves generating the refinement of the given boundary will be obtained by deletion of one of the creation operators and with the replacement with
an annihilation operator of the opposite vertex colour. 

\begin{figure}
\includegraphics[width=6cm]{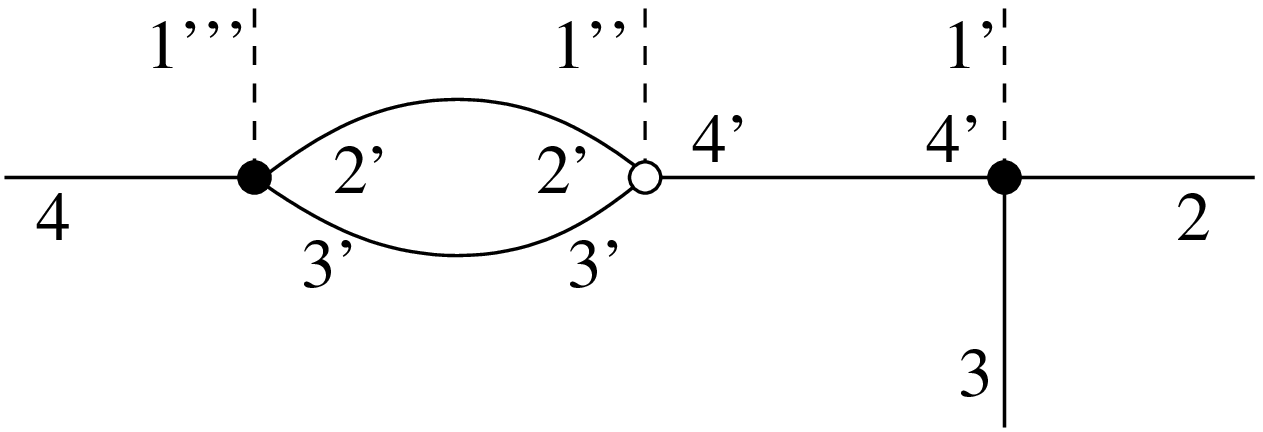}~~~~~~~~~
\includegraphics[width=6cm]{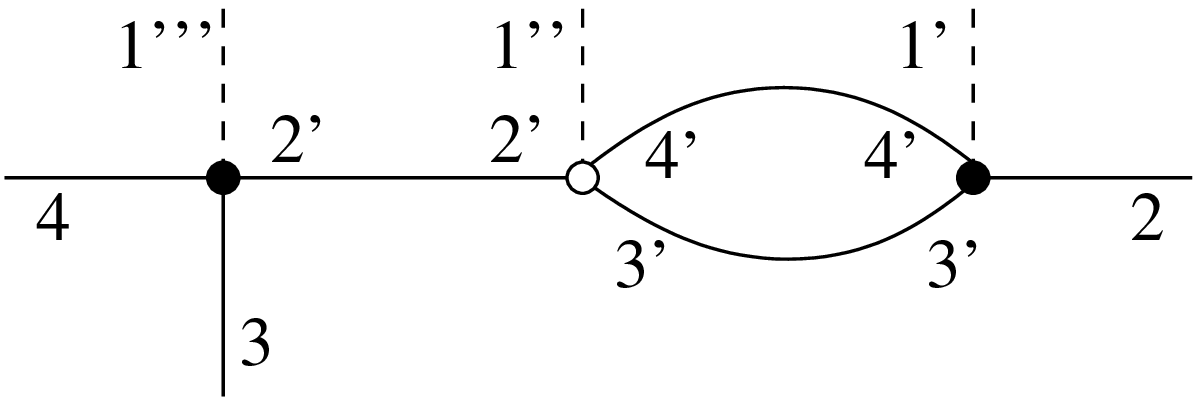}
\caption{The graph created by the move \eqref{move1} (left) and  \eqref{move2} (right). Notice that, besides the
legs of color 1 associated to the refinement of the boundary, the three open legs are
used for the gluing with the original graph.}
\label{figmove}
\end{figure}

The refinement involving the minimum number of vertices is generated, then, by two possible kinds of 1-3 move. 
More precisely, given for instance the boundary $-$ with color $1$ of the 
shell,
the refinement action of the corresponding operators consists into the replacement of a single black vertex with 3 new vertices
organized as in FIG. \ref{figmove}. By labeling the two move operators with the link colors  forming the dipole in the associated graphs and adopting the normal ordering, they read
\begin{align}
\move^{23}_{1,-}
\equiv \int 
&
 dk_2 dk_3 dk_4 dh_{4'} dh_{2'} dh_{3'} 
\nonumber
\\ 
&
\hsigma^{\dagger}_{\va B-}(e,k_{2},k_{3},h_{4'})
\hsigma^{\dagger}_{\va W-}(e,h_{2'},h_{3'},h_{4'})
\hsigma^{\dagger}_{\va B-}(e,h_{2'},h_{3'},k_{4})
\hsigma_{\va B-}(e,k_{2},k_{3},k_{4})
\label{move1}
\end{align}
and 
\begin{align}
\move^{43}_{1,-}
\equiv \int 
&
dk_2 dk_3 dk_4 dh_{4'} dh_{2'} dh_{3'} 
\nonumber
\\ 
&
\hsigma^{\dagger}_{\va B-}(e,k_{2},h_{3'},h_{4'})
\hsigma^{\dagger}_{\va W-}(e,h_{2'},h_{3'},h_{4'})
\hsigma^{\dagger}_{\va B-}(e,h_{2'},k_{3},k_{4})
\hsigma_{\va B-}(e,k_{2},k_{3},k_{4}).
\label{move2}
\end{align}

These operators preserve the topology of the graph and their explicit action can be deduced, as before, by inspection of the commutator:
\ba
&&[\move^{23}_{1,-},\hsigmad_{\va B-}(e,g_{2},g_{3},g_4) ]
=\n\\
&&\int  dk_2 dk_3 dk_4 dh_{4'} dh_{2'} dh_{3'} 
\hsigma^{\dagger}_{\va B-}(e,k_{2},k_{3},h_{4'})
\hsigma^{\dagger}_{\va W-}(e,h_{2'},h_{3'},h_{4'})
\hsigma^{\dagger}_{\va B-}(e,h_{2'},h_{3'},k_{4})
\times \n\\
&&[\hsigma_{\va B-}(e,k_{2},k_{3},k_{4}),
\hsigmad_{\va B-}(e,g_{2},g_{3},g_4) ]
=\n\\
 &&\int dh_{4'} dh_{2'} dh_{3'} 
\hsigma^{\dagger}_{\va B-}(e,g_{2},g_{3},h_{4'})
\hsigma^{\dagger}_{\va W-}(e,h_{2'},h_{3'},h_{4'})
\hsigma^{\dagger}_{\va B-}(e,h_{2'},h_{3'},g_{4}) 
\ea
and similarly for the operator \eqref{move2}. Therefore, when acting with the operator on a given
state, it will replace each vertex of the original state with a new subgraph, made with three
vertices, having three new open edges, while preserving the gluing

Analogous operators can be constructed to refine the $+$ boundary formed by open links with color 4 in FIG. \ref{trimelon}. It is easy to see that in this case the two operators would be labeled $\move^{12}_{4,+}, \move^{32}_{4,+}$. Extension to the case in which the boundaries $-$ and/or $+$ have a different color is immediate. Notice that for these simple refinement moves, in order to reproduce the four external links belonging to the original vertex in the right combinatorial order to be glued back to the rest of the graph, all the newly created radial links  need to have the same color as the original one; in this way, possible ambiguities resulting from links of different color forming the (refined) boundary are avoided. 

An important observation is that the sum of the fluxes $X$s along the four open links with color 1 in \eqref{boundary-ball} vanishes. This can be shown by using the gauge invariance \eqref{gauge-inv} at each node and the requirement that the parallel transported
 fluxes along common links between two nodes $v, v'$ to the same point (like, for instance, the middle of the link) satisfy $X_e(v)=-X_e(v')$, where $e$ is the color associated to the given common link. Such condition guarantees that the faces belonging to the tetrahedra $v,v'$ and dual to the common link can be correctly glued together so to allow to consistently reconstruct a given triangulation from the dual graph; this relation between the parallel transported fluxes can be seen to amount to requiring  the rotation of a flux induced by the group element associated to the corresponding link  to be on the plane orthogonal to the flux itself.
Then, the vanishing of the gravitational flux through the given boundary (in this case of color 1) can be written as
\be
X_1+X_{1'}+X_{1''}+X_{1'''}=0\,.
\ee
 In particular, this implies that the refinement moves \eqref{move1}, \eqref{move2} do not modify the total flux through the boundaries since
\be\la{flux-conservation}
X_1=X_{1'}+X_{1''}+X_{1'''}\,,
\ee
where $X_1$ is the flux along the initial link with color 1 in the boundary and $X_{1'}, X_{1''}, X_{1'''}$ the fluxes along the three new links, still with color 1, created in the refinement.

This minimal set of moves, needed to elucidate the main idea, has
to be applied in all the possible ways to the seed triangulation encoded in the initial state \eqref{tau}.
Repeated action of these operators
leads to the growth of the seed state into states associated to finer and finer triangulations.
In particular, the (not normalized) state
\begin{equation}
\exp\left( \move_{tot} \right) \ket{\tau}, \qquad  \move_{tot}= 
\move^{23}_{1,-}+
\move^{43}_{1,-}+
\move^{12}_{4,+}+
\move^{32}_{4,+}
+\move_{\va W0}+ \move_{\va B0}
\end{equation}
does represent a superposition of condensed vertices glued to form a single spherically
symmetric shell in all the possible ways compatible with the action of the set of basic
moves that we have described above. It is a particular choice that keeps to a minimum
the additional parameters controlling the state, essentially fixing the coefficients
of the linear combination of different triangulations appearing in its decomposition into
the components with fixed number of particles. Other choices are certainly possible.

The inclusion of a larger set of moves, once
they have been constructed and classified, will be straightforward. 

We emphasise again that all the moves
involve field operators that are associated to the same
wave-function. 
Therefore, the total state is still controlled by the single vertex wave-functions associated
to the original state $\ket{\Omega}$ that we are starting from. No additional parameters
are added, and the state can still be seen as a condensate of spin network vertices.

We also stress that, given the definitions above, the state does not single out any preferred
triangulation. In particular, given that it can be decomposed in terms of states
with an arbitrarily large number of vertices, it might be interpreted as defining a kinematical continuum limit for quantum geometries.

\subsubsection{The ball}

A particular example of shell is given by a 3-ball obtained from the graph in FIG. \ref{trimelon} by gluing together the two open links forming one of the two boundaries, while keeping the other two open. For instance, if we glue the two link of color 1 together, the seed state for the 3-ball can be written as
\ba\la{B}
\ket{B^3} &=&
\int (dg)^{11} 
\hsigmad_{\va B0}(g_1,g_{2},g_{3},g_4)
\hsigmad_{\va  W0}(g_1,g_{2'},g_{3},g_4)
\hsigmad_{\va  B0}(g_{1''},g_{2'},g_{3'''},g_{4''})\n\\
&&~~~~~~~~~~~\hsigmad_{\va  W0}(g_{1''},g_{2},g_{3''},g_{4''})
\hsigmad_{\va B+}(g_{1''''},g_{2''''},g_{3''},e)
\hsigmad_{\va  W+}(g_{1''''},g_{2''''},g_{3'''},e)
\fv \, ,
\ea
where we have replaced the label $-$ with $0$ for the two field operators which no longer belong to boundary.

The refinement then proceeds like in the general case of the shell, with the operator $\move_{\va W0}$ in \eqref{W0} and its analog $\move_{\va B0}$ for the bulk vertices and the operators $\move^{12}_{4,+}, \move^{32}_{4,+}$ for the vertices in the only boundary left.

%

%%%%%%%%%%%%%%%%%%%%%%%%%%%%%%%%%%%%%%%%%
%%%%%%%%%%%%%%%%%%%%%%%%%%%%%%%%%%%%%%%%%
\section{States and expectation values}\la{sec:vev}

Having defined the context and the technical tools required, we can then start to
explore the space of states that can be generated in this way. The most general
states that we can build combining the ideas introduced so far are of the form
\begin{equation}
\ket{\Psi_F}=F(\move) \ket{seed}
\end{equation}
and are parametrised by the function $F$ and the wave-function $\sigma$. While the
kind of graphs that are generated is encoded essentially in the seed and in the
structure of the refinement move,  
the function
$F$ has the role of specifying which kind of superposition of graphs is contained in the state,
and the coefficient of the linear combination, as determined by the coefficients of the
Taylor series of $F$.
For the sake of simplicity, and in analogy with coherent states, we can restrict the function $F$ to be of exponential form. In general, the choice of the function $F$ can be motivated either on the basis of simplicity or
in terms of some physical arguments, e.g like the maximum entropy argument spelled out in \cite{fidelity}.

Notice that a single monomial $\move^n$, when acting on a given state, does not generate
a single state associated to a given graph, but rather a superposition of states, each of which
labeled by one of the graphs obtained acting with a particular sequence of refinement moves,
in all the possible ways.

We now address two interesting points: normalisation of the states and
expectation values of one-body operators.

If we work with the restriction \eqref{restriction}, it is straightforward to realize that the
seed states and all the ones generated from them via refinement moves are not normalisable.
However, they are not dissimilar from the ones representing eigenstates of the holonomy
operators. Indeed, if one considers the effects of \eqref{restriction} on the definition
of $\hsigma$, one sees that, apart from the insertion of arbitrary phase factors which are not
determined by the commutation relation,
the operator is the mirror image of $\hphi$, in a Peter--Weyl decomposition, 
with an exchange of left and right indices.

Consequently, one can expect that they can be seen as limit of sequences of
normalisable states.
The computation of generic expectation values, then, should proceed 
with a suitable regularization of the expression, followed by the appropriate limit.
This is implicitly assumed in what follows. 

Much can be learned, even neglecting this issue, from the inspection of matrix elements of simple observables
represented by one body operators\footnote{As we have already mentioned, the analysis of observables associated to a larger number of particles presents additional interpretational
challenges which obscure the significance of the mathematical results. We will ignore them,
in this paper, as they play no role in what we want to say.}.

Given a ``first-quantized'' observable $A$, as it might be defined in an LQG-like context (\eg
the volume operator, the flux dual to an edge, etc.), we represent its matrix elements
with a kernel:
\begin{equation}
A(g_v,g_w) \equiv ( g_v | \hat{A} | g_w )
\end{equation}
We then promote it to a second quantized operator in the familiar way \cite{GFT-LQG}:
\begin{equation}
\widehat{\mathcal{A}} \equiv \int \extd g_v \extd g_w A(g_v,g_w) \hphid(g_v) \hphi(g_w).
\end{equation}

We first consider the particular class of one body operators for which $\sigma$ is an eigenfunction
\begin{equation}
(A\rhd\overline{\sigma})(g_v) = \int \extd g_w A(g_v,g_w) \overline{\sigma(g_w)} =a_{\sigma} \overline{\sigma(g_v)},
\end{equation}
For these states:
\begin{align}
& [\widehat{\mathcal{A}}, \hsigmad(h_t)]= 
\int \extd g_v \extd g_w  \extd g_u 
A(g_v,g_w)
\overline{\sigma(h_t g_u)}
[\hphid(g_v) \hphi(g_w),\hphid(g_u)]= \n
\\&
\int \extd g_v \extd g_w  
A(g_v,g_w)
\overline{\sigma(h_t g_w)}\hphid(g_v) = 
\int  \extd g_v \left( A \rhd T_{h_{t}} \rhd \overline{\sigma} \right)(g_v)
\hphid(g_v) 
\end{align}
where $T_{h_t}$ is the left translation on the group by the group elements $h_t$.

If we restrict the attention to the operators such that $
[A,T_h] = 0
$, 
then:
\begin{equation}
[\widehat{\mathcal{A}}, \hsigmad(h)] = a_{\sigma} \hsigmad(h)
\end{equation}
and, as a consequence, the expectation value of the one body operator on the glued
state is simply
\begin{equation}
\langle \widehat{\mathcal{A}} \rangle = a_{\sigma} \langle \widehat{N} \rangle
\end{equation}
for any pattern of gluings.

In general, however, $
[A,T_h] \neq 0
$ {and the computation of expectation values requires an extended discussion.
Define for short
\begin{equation}
\hat{\alpha}^{\dagger}(h) = [\widehat{\mathcal{A}}, \hsigmad(h)]  =
\int  \extd g_v \left( A \rhd T_{h} \rhd \sigma \right)(g_v)
\hphid(g_v) \,
\end{equation}
and consider the expectation values for a fixed graph
\begin{align}
&\bra{\Gamma,\sigma} \widehat{\mathcal{A}} \ket{\Gamma,\sigma} =
\sum_{v \in \Gamma} \bra{\Gamma,\sigma}\left(\int dh \hat{\alpha}^{\dagger}(h) \ket{\Gamma\setminus v, h,\sigma}
\right)\, .
\end{align}
As the action of the operator $\hat{\mathcal{A}}$ changes the wave-function in a noncontrolled way, we seem to be at a dead end. Notice, however, that as the
states are completely symmetrized (an aspect of the refinement procedure we implemented), the expectation value scales with the number of
vertices $N$:
\begin{align}
\sum_{v \in \Gamma} \bra{\Gamma,\sigma}\left(\int dh \hat{\alpha}^{\dagger}(h) \ket{\Gamma\setminus v, h,\sigma}
\right) = N(\Gamma)
\bra{\Gamma,\sigma}\left(\int dh \hat{\alpha}^{\dagger}(h) \ket{\Gamma\setminus v, h,\sigma}
\right) = N(\Gamma) a(\Gamma) \, ,
\end{align}
where with $a(\Gamma)$ we denote a quantity that is intensive, and not extensive, which might
differ from $a_{\sigma}$ due to the appearance of the connectivity  in the definition of
the state.

Let us examine this point in greater detail.
A state belonging to the linear combination $F(\move)\ket{seed}$ will be built with the convolution of a connectivity kernel $J_{\Gamma}$ and the field operators:
\begin{equation}
\ket{\Gamma_n}=\int \prod_{i=1}^n \extd h_{\va v_i} \, J_{\Gamma}(h_{\va v_1},\ldots,h_{\va v_n}) \hsigmad(h_{\va v_1})\cdots \hsigmad(h_{\va v_n}) \ket{0}.
\end{equation}
This $J_{\Gamma}$ is made by a number of Dirac deltas, identifying the pairs
of group elements that are used to form a link via a convolution. As our fields are all identical,
the kernel is completely symmetrized. In a case in which additional labels are present (e.g. matter
fields, other colours), this might not be the case,  and the kernel will be completely simmetrized only in those slots referring to the same labels.

Consider now
the matrix element
\begin{align*}
\bra{\Gamma_n}\widehat{\mathcal{A}} \ket{\Gamma_n} =& \int \prod_{i=1}^n \extd h_{\va v_i}  \extd h'_{\va v_i} \extd g_v \extd g_w\,
J_{\Gamma}(h_{\va v_1},\ldots,h_{\va v_n})
 J_{\Gamma}(h'_{\va v_1},\ldots,h'_{\va v_n}) A(g_v,g_w)\times
 \n\\
 &
 \bra{0} 
 \left[\hsigma(h'_{\va v_1})\ldots \hsigma(h'_{\va v_n}),\hphid(g_v) \right]
 \left[\hphi(g_w),\hsigmad(h_{\va v_1})\ldots \hsigmad(h_{\va v_n}) \right] \ket{0}
\n\\=&
 \,n^2
 \int \prod_{i=1}^n \extd h_{\va v_i}  \extd h'_{\va v_i} \extd g_v \extd g_w
J_{\Gamma}(h_{\va v_1},\ldots,h_{\va v_n})
 J_{\Gamma}(h'_{\va v_1},\ldots,h'_{\va v_n}) A(g_v,g_w)
 \sigma(h_{\va v_1}'g_v)
 \overline{\sigma(h_{\va v_1} g_w)} 
\n \\
 &
 \times\bra{0} 
 \hsigma(h'_{\va v_2})\ldots \hsigma(h'_{\va v_n})
 \hsigmad(h_{\va v_2})\ldots \hsigmad(h_{\va v_n})\ket{0}\,.
\end{align*}
Keeping in mind the Wick's theorem, we count $(n-1)!$ possible contractions, and we obtain
as a consequence:
\be
\bra{\Gamma_n} \!\widehat{\mathcal{A}}  \ket{\Gamma_n}\! =\! n \cdot n!\!\! \int\!\! \prod_{i=1}^n \extd h_{\va v_i}  \extd h'_{\va v_i}  \extd g_v \extd g_w
J_{\Gamma}(h_{\va v_1},h_{\va v_2},\ldots,h_{\va v_n})
 J_{\Gamma}(h'_{\va v_1},h_{\va v_2},\ldots,h_{\va v_n}) 
 A(g_v,g_w)\sigma(h_{\va v_1}'g_v)
 \overline{\sigma(h_{\va v_1} g_w)};
\ee
we can further massage this expression, with the following definition:
\begin{align}
&\int \extd h_{\va v_2} \ldots  \extd h_{\va v_n} J_{\Gamma}(k_{\va v_1},h_{\va v_2},\ldots,h_{\va v_n})
 J_{\Gamma}(k'_{\va v_1},h_{\va v_2},\ldots,h_{\va v_n}) \n\\
 &= C_\Gamma(k_{\va v_1},k'_{\va v_1})  \int  \extd h_{\va v_1} \ldots \extd h_{\va v_n} J_{\Gamma}(h_{\va v_1},h_{\va v_2},\ldots,h_{\va v_n})
 J_{\Gamma}(h_{\va v_1},h_{\va v_2},\ldots,h_{\va v_n})
\end{align}
which leads to
\begin{equation}
\frac{\bra{\Gamma_n}\widehat{\mathcal{A}}  \ket{\Gamma_n}}{\braket{\Gamma_n}{\Gamma_n}}
= n \int \extd h_{\va v_1}  \extd h'_{\va v_1} \extd g_v \extd g_w 
 C_\Gamma(h_{\va v_1},h'_{\va v_1})
 A(g_v,g_w)\sigma(h_{\va v_1}'g_v)
 \overline{\sigma(h_{\va v_1} g_w)}\,.
\end{equation}
This is yet another way to show that, in principle, the connectivity of the graph influences the expectation values of a one body operator. 

However, if the graph is obtained starting from a carefully chosen initial state by carefully designed refinement moves (which is the case for the constructions presented in Sections \ref{sec:operators}, \ref{sec:sphere} and  \ref{sec:shells}), this kernel $C_\Gamma$ can be made independent from $\Gamma$, and in fact can be taken to be $\delta(h_{\va v_1},h_{\va v_1}')$.

This can be seen as follows: the result of the integrations with the $\delta$s encoding the graph structure, apart from possibly diverging coefficients, has the form, for each possible pairing of
operators:
\begin{equation}
\int \Delta_L(h_{v_1},\pi h'_{v_1})
 A(g_{v},g_{w})\sigma(h'_{v_1} g_{v})\overline{\sigma(h_{v_1} g_{w})} \extd h_{\va v_1} 
 \extd h'_{\va v_1} \extd g_v \extd g_w.
\end{equation}
where $\pi$ is a permutation of the legs of the vertex that depends on the pairing. If one
adds colours, it is immediate to see that $\pi$ has to be the identity. However, in the
case in which $\hsigma(\pi g) = \hsigma(g) \qquad \forall \pi$ (a condition that seems to
be implicit if \eqref{restriction} holds) one sees that the permutation $\pi$ can be trivialized.

Therefore, in this case, the expression is common to all the terms in the sum, and
\begin{equation}\la{expectation}
\langle \widehat{\mathcal{A}}  \rangle = \langle \widehat{\mathcal{N}}  \rangle
\int \extd h_v \extd g_v \extd g_w
\sigma(h_{v} g_{v}) A(g_{v},g_{w})\overline{\sigma(h_{v} g_{w})},
\end{equation}
which is the expected result, highlighting not only the extensive nature of the one
body operator, but also the impact of the gluing.

The expression so obtained is very general. Even if restricted to one-body operators, it already covers most observables that are useful in a (quantum) cosmological context \cite{GO}. As an example of a useful operator, let us now compute the expectation value of the number operator (the simplest one-body operator)  for one of the homogenous states constructed above, namely an arbitrarily refined 3-sphere
\be\la{3-sphereR}
|S^3_{\va R}\rangle = e^{\move_{\va W}+\move_{\va B}}|S^3\rangle\,,
\ee
where the 3-sphere seed state $|S^3\rangle$ is given in \eqref{seed-sphere}, the refinement operators by \eqref{move-circ} (and its analog for a black vertex). 

A similar calculation applies to the simplicial case and to the other topologies presented above.

\subsubsection{Number Operator}\la{NO}

In the case of a 3-sphere state constructed out of melonic graphs, the number operator for both black and white vertices is defined by
\be
\numop=\numop_{\va B}+\numop_{\va W}=\int \extd g_v \left (\hphid_{\va B} (g_v) \hphi_{\va B} (g_v)+\hphid_{\va W} (g_v) \hphi_{\va W} (g_v)\right)\,.
\ee
It is immediate to see that $\numop$ acts diagonally on a given vertex state, namely
\ba\la{N}
\numop_{\va B} \hsigmad_{\va B}(h_w)|0\rangle&=& [\numop_{\va B}, \hsigmad_{\va B}(h_w)]|0\rangle\n\\
&=& \int \extd g_v\, \hphid_{\va B} (g_v)  \int  \extd k_w\, \sigma(h_w k_w)[\hphi_{\va B} (g_v), \hphid_{\va B} (k_w)]|0\rangle\n\\
&=& \int \extd g_v\, \hphid_{\va B} (g_v)  \int  \extd k_w\int d\gamma \, \sigma(h_w k_w) \prod_{i=1}^4 \delta ( g_{(v,i)} \gamma k^{-1}_{(w,i)})|0\rangle \n\\
&=& \int \extd g_v\int d\gamma\,  \sigma(h_w g_v\gamma)  \hphid_{\va B} (g_v) |0\rangle\n\\
&=& \hsigmad_{\va B}(h_w)|0\rangle\,,
\ea
where we used the canonical commutation relations  \eqref{CCRright} and  the gauge invariance of the wave-function; a similar calculation shows 
\ba
&&\numop_{\va W}\hsigmad_{\va W}(h_w)|0\rangle = \hsigmad_{\va W}(h_w)|0\rangle\\
&&\numop_{\va B}\hsigmad_{\va W}(h_w)|0\rangle=0= \numop_{\va W}\hsigmad_{\va B}(h_w)|0\rangle\,.
\ea

We can now compute the expectation value
\ba\la{Nexp}
\langle S_{\va R}^3 |\numop|S_{\va R}^3\rangle &=& \langle S^3 |  e^{\moved_{\va W}+\moved_{\va B}} (\numop_{\va W} + \numop_{\va B}) e^{\move_{\va W}+\move_{\va B}}|S^3\rangle \n\\
&=& \langle S^3 |  e^{\moved_{\va B}} e^{\moved_{\va W}}\numop_{\va W}  e^{\move_{\va W}}e^{\move_{\va B}}|S^3\rangle+  \langle S^3 |  e^{\moved_{\va B}}  e^{\moved_{\va W}} \numop_{\va B} e^{\move_{\va W}}e^{\move_{\va B}}|S^3\rangle\,,
\ea
where in the last passage we used the commutativity of the two refinement operators $\move_{\va W}, \move_{\va B}$, as it can easily be seen by the action of the second operator on one of the new vertices created by the action of the first (the action on different vertices of the original graph commutes trivially). Let us now concentrate on the first of the two terms in the last line on \eqref{Nexp}, since the second expectation value gives exactly the same result due to the same number of new black and white vertices created by any of the two refinement operators at each step (and the symmetry in black and white vertices of the seed state, of course). Let us start by computing 
\baa
\numop_{\va W}  e^{\move_{\va W}}e^{\move_{\va B}}|S^3\rangle&=&
\numop_{\va W}\sum_{n_{\va W}, n_{\va B} =0}^\infty \frac{1}{n_{\va W}!}\frac{1}{n_{\va B}!} (\move_{\va W})^{n_{\va W}}(\move_{\va B})^{n_{\va B}}|S^3\rangle\\
&=&\numop_{\va W}\sum_{n_{\va W}, n_{\va B} =0}^\infty \frac{(n_{\va W}+n_{\va B}+1)!}{n_{\va W}!n_{\va B}!}
\big| \underbrace{\begin{array}{ccc}
\van \circ & \van\cdots & \van \circ \\
\van \bullet & \van\cdots &\van  \bullet \end{array} }_{n_{\va W}+n_{\va B}+2}\big\rangle\\
&=&\sum_{n_{\va W}, n_{\va B} =0}^\infty \frac{(n_{\va W}+n_{\va B}+2)!}{n_{\va W}!n_{\va B}!}
\big| \underbrace{\begin{array}{ccc}
\van \circ & \van\cdots & \van \circ \\
\van \bullet & \van\cdots &\van  \bullet \end{array} }_{n_{\va W}+n_{\va B}+2}\big\rangle\,,
\eaa
where the state in the last two lines represents a refined 3-sphere containing $n_{\va W}+n_{\va B}+2$ black and white vertices. If we now act with the adjoint refinement operators $\moved_{\va W}, \moved_{\va B}$ on such state to remove vertices according to the inverse of the move depicted in \eqref{move-circ} (and its analog for a black vertex refinement), we obtain the expectation value
\ba
\langle S_{\va R}^3 |\numop_{\va W}|S_{\va R}^3\rangle &=&
 \sum_{n_{\va W}, n_{\va B} =0}^\infty
\sum_{n'_{\va W} =0}^{ n_{\va W}+ n_{\va B}+1}\delta_{n'_{\va B} , n_{\va W}+ n_{\va B}+1-n'_{\va W}}\frac{1}{n_{\va W}!n_{\va B}!n'_{\va W}!n'_{\va B}!}\n\\
 &\times& \left[\frac{(n_{\va W}+ n_{\va B}+2-n'_{\va W})!}{(n_{\va W}+ n_{\va B}+2-n'_{\va W}-n'_{\va B})!}\right]^2
 \frac{(n_{\va W}+ n_{\va B}+1-n'_{\va W})!}{(n_{\va W}+ n_{\va B}+1-n'_{\va W}-n'_{\va B})!}\n\\ 
&\times& \left[ \frac{(n_{\va W}+ n_{\va B}+2)!}{(n_{\va W}+ n_{\va B}+2-n'_{\va W})!}\right]^2
\frac{(n_{\va W}+ n_{\va B}+1)!}{(n_{\va W}+ n_{\va B}+1-n'_{\va W})!}\n\\
  &\times&(n_{\va W}+n_{\va B}+2)!\big\langle \begin{array}{ccc}
\van \circ\\ \van \bullet \end{array}  \big| \begin{array}{ccc}
\van \circ\\ \van \bullet \end{array}  \big\rangle\n\\
&=& \sum_{n_{\va W}, n_{\va B} =0}^\infty
\sum_{n'_{\va W} =0}^{ n_{\va W}+ n_{\va B}+1} \frac{[(n_{\va W}+n_{\va B}+2)!]^2(n_{\va W}+n_{\va B}+1)!}{n_{\va W}!n_{\va B}!n'_{\va W}!(n_{\va W}+ n_{\va B}+1-n'_{\va W})!}\n\\
 &\times&(n_{\va W}+n_{\va B}+2)!\big\langle \begin{array}{ccc}
\van \circ\\ \van \bullet \end{array}  \big| \begin{array}{ccc}
\van \circ\\ \van \bullet \end{array}  \big\rangle
\ea
and the same expression is obtained for $\langle S_{\va R}^3 |\numop_{\va B}|S_{\va R}^3\rangle $. The norm of the dipole state $ \big| \begin{array}{ccc} \van \circ\\ \van \bullet \end{array}  \big\rangle$ is given by
\ba
\big\langle \begin{array}{ccc}
\van \circ\\ \van \bullet \end{array}  \big| \begin{array}{ccc}
\van \circ\\ \van \bullet \end{array}  \big\rangle&=&
\int  \extd g_v \extd h_v \langle 0|    \hsigma_{\va W}(h_v) \hsigma_{\va B}(h_v) \hsigmad_{\va W}(g_v)  \hsigmad_{\va B}(g_v)|0\rangle\n\\
&=&\int \extd g_v \extd h_v \extd k_w  \extd k'_v \extd\tilde k_v \extd\tilde k'_v \overline{\sigma(h_v k_v)}\, \overline{\sigma(h_v k'_v)}
 \sigma(g_v \tilde k_v) \sigma(g_v \tilde k'_v) \n\\
 &\times& \langle 0| [\hphi_{\va W}(k_v),\hphid_{\va W}(\tilde k_v)]  [\hphi_{\va B}( k'_v),\hphid_{\va B}(\tilde k'_v)] |0\rangle\n\\
 &=&\int \extd g_v \extd h_v \extd k_v \extd k'_v  \overline{\sigma(h_v k_v)}\, \overline{\sigma(h_v k'_v)}
 \sigma(g_v  k_v) \sigma(g_v  k'_v) \,,
\ea
where in the last passage we used the commutation relations \eqref{CCRright} and the wave-function gauge invariance \eqref{left-right-inv}. For the special class of wave-function $\sigma$ found in Appendix \ref{app:calculationsrefinement} such that eq. \eqref{c-restriction} is satisfied, the expression above for the norm of the dipole state is not normalizable (see also the discussion at the beginning of the previous section).
However, it can be seen that the normalized expectation value $\langle S_{\va R}^3 |\numop|S_{\va R}^3\rangle/\langle S_{\va R}^3 |S_{\va R}^3\rangle$ is well defined and grows linearly in $n_{\va W}+ n_{\va B}$ \footnote{Let us mention that most of the above, with the exclusion of issues deriving from the combinatorial structure of our quantum states, but including the issues with normalising condensate states like ours in the scalar product of the initial Fock space, is simply the GFT counterpart of standard issues in QFT, in their condensed matter theory applications.}.

%%%%%%%%%%%%%%%%%%%%%%%%%%%%%%%%%%%%%%%%%
%%%%%%%%%%%%%%%%%%%%%%%%%%%%%%%%%%%%%%%%%

\section{Geometric data for Spherical Shells}\label{sec:geometry}

In order to extract the geometric information out of the states that we have designed, we need to translate into the appropriate second quantised form the
geometric operators of LQG. To keep the discussion concrete, we consider, for the moment, just the flux and area operators. Indeed, with these operators
we can reconstruct the part of the intrinsic geometry of the shell related to the area of the boundaries, \ie, in the case of closed graphs, the function $R^2(r)$.

From the very construction of the states as condensed states, we can argue that they are indeed associated to spherically symmetric
geometries, as the geometry itself, stored by the wave-functions associated to the vertices, is the same at each vertex (and this fact is preserved by
the refinement). Therefore, our shells are homogeneous
in the angular directions.

The only thing that has to be explicitly checked is that the surfaces bounding the shells are indeed closed surfaces. In order to do so, we need
to check that the fluxes associated to the radial links add up to zero.

It is immediate to give a local definition of the second quantised version of the flux operator, acting on each single vertex and computed on a face\footnote{In this section we use a slightly different notation: instead of $g_v\equiv g_{(v,i)}, i=1,2,3,4$, to indicate the four group elements argument of the field operators, we use $g_I, I=1,2,3,4$; similarly, the Haar measure is now indicated by $dg_I\equiv dg_1dg_2dg_3dg_4$.}
dual to the $J-$th link:
\begin{equation}
\EE^{i}_{J,cs} \equiv \int dg_I \hphid_{cs}(g_I) E^{i}_{J} \rhd \hphi_{cs}(g_I)
= \int dg_I E^{i}_{J}\rhd \hphid_{cs}(g_I)\hphi_{cs}(g_I)
\end{equation}
where we used the definition
\begin{equation}
E^{i}_{J} \rhd f(g_I) := \lim_{\epsilon\rightarrow 0} \iu \frac{d}{d\epsilon}
f(g_{1},\ldots, e^{-\iu \epsilon \tau^{i}}g_{J}, \ldots, g_{4})
\end{equation}
and the fact that this operator is Hermitian on $\mathcal{L}^2(G)$.

It is immediate to compute the action on the states that we are interested in, once we compute the commutator:
\begin{align}\la{E-expectation}
\left[\EE^{i}_{J,cs}, \hsigmad_{c's'}(h_I)  \right] & =
\delta_{cc'}\delta_{ss'} \int dg_I
\left[ \hphid_{cs}(g_I) E^{i}_{J} \rhd \hphi_{cs}(g_I), \hsigmad_{cs}(h_I)\right]
 \nonumber \\
& =\delta_{cc'}\delta_{ss'} \int dg_I\int dk_I
 \overline{\sigma(h_I k_I)}
 \left[ \hphid_{cs}(g_I) E^{i}_{J} \rhd \hphi_{cs}(g_I), \hphid_{cs}(k_I)\right]  \nonumber \\
 &= \delta_{cc'}\delta_{ss'} \int dg_I\int dk_I\int d\gamma \overline{\sigma(h_I k_I)}
 E^{i}_{J}\rhd \hphid_{cs}(g_I) )\prod_{I=1}^4 \delta ( g_I \gamma k^{-1}_I)  \nonumber \\
 & =\delta_{cc'}\delta_{ss'} \int dg_I 
 E^{i}_{g_J}\rhd \overline{\sigma(h_I g_I)}
 \hphid_{cs}(g_I)
\end{align}
where we are using repeatedly the Hermiticity of the flux operator and again the commutator  \eqref{CCRright} together with gauge invariance of the wave-function.

In this way we have defined a family of flux operators which are acting colour by colour. This means that
the flux operator $\EE^{i}_{1,B+}$ will measure the sum of all the fluxes of the black vertices of the boundary $+$, crossed
by edges of colour $1$. All the other edges of colour $1$ appearing in the graph will be correctly ignored\footnote{As we mentioned above, the labelling we are using is probably meant to represent a more complicated procedure of reconstruction
of the geometry and topology associated to the states used, and that the percolation of these label into the second quantised
observables has to be intended in a similar fashion.}. 

The expectation value of the flux operator on a generic fixed graph (obtained for instance at a given stage of the refinement process described above) can then be computed from \eqref{E-expectation} and the general expression \eqref{expectation}, being $\EE$ a one-body operator.

The conditions of closure of the initial state, then, can be written as:
\begin{equation}
\left(\EE^{i}_{J,Bs}+\EE^{i}_{J,Ws}\right) \ket{\{\tau\}} = 0.
\end{equation}
If this closure relation holds, i.e. if the fluxes through the two boundaries of the shell vanish separately, for the initial state then it holds also for the state at any stage of refinement. That this is the case, namely that
\begin{equation}
\left(\EE^{i}_{J,Bs}+\EE^{i}_{J,Ws}\right) \exp(\allmoves) \ket{\{\tau\}} =\exp(\allmoves)  \left(\EE^{i}_{J,Bs}+\EE^{i}_{J,Ws}\right) \ket{\{\tau\}}=0\,,
\end{equation}
follows from the property \eqref{flux-conservation} of our refinement moves.

Following a similar path, one can obtain the total area operator(s):
\begin{equation}
\hat{\mathbb{A}}_{J,cs} \equiv \int dg_I \hsigmad_{cs}(g_I) \sqrt{E^{i}_{J} E^{j}_{J} \delta_{ij}} \rhd \hsigma_{cs}(g_I)
\end{equation}
which is decorated with the necessary labels required to concentrate the action on the relevant portion of the state, thus counting only the contribution of the dual plaquettes that we are considering. 

The procedure described so far does not really depend on the details of the operator
that we are considering, apart from it being a one-body operator. Therefore, we can
extend the construction given above to arbitrary one-body operators  as we
have already seen.
If we want then to compute the contribution to the desired observables coming from a specific
region of the graph, then, we can simply replace the appropriate $\hsigma_{cs},\hsigmad_{cs}$
in the expression for $\widehat{\mathcal{A}}$. This might be useful if we were looking, for example, at properties like the total volume of the shell.

%%%%%%%%%%%%%%%%%%%%%%%%%%%%%%%%%%%%%%%%%
%%%%%%%%%%%%%%%%%%%%%%%%%%%%%%%%%%%%%%%%%

\section{Conclusions and a brief outlook}\la{sec:conclusions}

We have generalised the construction of quantum gravity condensate states performed in \cite{cosmoshort,cosmolong} as candidates for representing continuum homogeneous quantum geometries, and to extract effective cosmological dynamics from the fundamental quantum gravity dynamics. Indeed, these states are (kinematical) quantum states of the full GFT/LQG Hilbert space, not resulting from any classical symmetry reduction. They correspond to homogeneous continuum geometries in that, while defined in terms of an infinite superposition of spin network states defined on arbitrarily fine graphs, are collectively described by a single wave function for minisuperspace variables. They do realise, therefore, a coarse graining of (infinite) fundamental quantum gravity degrees of freedom to cosmological variables, at least at the kinematical level. 

These states depend on only three structures:
\begin{enumerate}
\item the seed state $\ket{\tau}$;
\item the single vertex wave-function $\sigma$;
\item the combinatorics of the move operator $\move$.
\end{enumerate}
As such, they are not unique and one can conceive easily different states. Still, they are rather simple and amenable to manipulations and generalizations. 

An important feature of these new states is that they contain the basic connectivity information required to reconstruct the topological structure of the desired geometry. This is crucial for generalisations to more involved (less symmetric) quantum geometries. In order to achieve the needed control over the combinatorial structures associated to these states, we have shown that tools from tensor models are essential.

Our construction also illustrates how the Fock structure behind (the reformulation of LQG as) GFT allows a straightforward definition of refinement operators for spin network states. Such type of operators, we believe, can have many applications in quantum gravity, beyond our specific construction (tied to the specific notion of ``homogeneity'' we have argued for). 

One obvious direction for future work is to clarify under which mathematical conditions the states we have constructed reduce to the simpler condensate states used in  \cite{cosmoshort,cosmolong} and to elucidate the physical meaning of those conditions. 

In the opposite direction, one can instead use the new condensate states we have constructed as building blocks for constructing quantum states associated to less symmetric continuum quantum geometries, still within the full theory.

The natural next stage in a process of ``quantum spacetime engineering''  is to construct quantum states associated to continuum quantum geometries endowed with a notion of spherical symmetry. These would be the framework to study, for example, quantum black holes within the fundamental theory, rather than in simplified models inspired by it, as we have started to do for cosmology. 

This is going to be subject of future work, but the main idea is the following. Having constructed quantum states associated to homogeneous 3-shells, and to homogeneous 3-balls, one can then glue such homogeneous 3-shells into a complete foliation. The data associated to each shell (better, to its radial open legs) will also have to be compatible with spherical symmetry. This requires two things. First, the wave function associated to each shell has to be different, and the set of such wave functions can be parametrised by a single label, effectively playing the role of radial coordinate. Second, for each face (inner and outer) of each shell, the wave-function has to be chosen in such a way that  the spins are composed to give the representation of spin-0 (this is the algebraic encoding of spherical symmetry). In other words, one has to be able to contract all the magnetic indices of the $\SU(2)$ matrices associated to each radial link with a single big intertwiner, i.e. if the graph on each shell can be corse-grained in a gauge invariant way to a single vertex, then spherical symmetry is guaranteed. For this to be possible, one simply requires that the graph on each shell be connected. 
In fact, one can use, for instance, techniques introduced in \cite{Flower} to map by a gauge transformation all the group elements associated to the links forming a maximal tree of the shell angular graph to the identity. In this way, the shell wave-function is gauge-fixed to a flower combinatorial structure formed by a single vertex, from which all the radial links depart, and as many petals (loops) as the number of shell original  angular links minus the number of vertices plus one. The only non-trivial holonomies left are only those along the petals and there is a global single $\SU(2)$ gauge invariance remaining on each shell and encoding its spherical symmetry. This construction should also allow a direct comparison with the simple states used in \cite{martin, eteradanny} for modelling quantum black holes. Finally, a proper study of black holes within the full theory, using such ``spherically symmetric multi-condensate states'' will also require the imposition of isolated-horizon conditions \cite{IH} on them (better, on their outer boundaries); moreover, our construction is well suited to the definition of Kubo-Martin-Schwinger states along the lines of   \cite{KMS}. Preliminary results of this analysis have been reported in \cite{BH-condensate}.

%%%%%%%%%%%%%%%%%%%%%%%%%%%%%%%%%%%%%%%%%
%%%%%%%%%%%%%%%%%%%%%%%%%%%%%%%%%%%%%%%%%

\appendix

\section{An example for the refinement move}\label{app:calculationsrefinement}
\subsection{The wave-function}

For our refinement moves introduced in the paper to satisfy the desired property \eqref{sim-move}, we found a solution given by the condition \eqref{restriction}. Therefore, we now investigate which restrictions on the wave-function $\sigma$ such condition implies, reducing the family of states allowed by our construction. Let us point out again that these are sufficient conditions for having  \eqref{sim-move}, not necessary ones, and the only purpose here is to show that the construction is not empty, as finding general solutions is difficult.

We start with the examination of the wave-function. As we ask gauge invariance from the left and the right, the Peter--Weyl decomposition gives simply:
\begin{equation}
\sigma(g_v) = \sum_{\{j\},l_L,l_R} 
{\suinter{1}{2}{3}{4}{l_L}{m}
\suinter{1}{2}{3}{4}{l_R}{n}}
\sigma^{j_1j_2j_3j_4 l_L l_R} 
\prod_{i=1}^{4} D^{j_{i}}_{m_{i}n_i}(g_{(v,i)}) \, .
\end{equation}
Our conventions are:
\begin{equation}
D^{j}_{mn}(g)D^{j'}_{m'n'}(g) = \sum_{j'' \geq |j-j'|}^{j+j'} 
C^{j j' j''}_{m m' m''}
C^{j j' j''}_{n n' n''}
D^{j''}_{m''n''}(g)
\end{equation}
and
\begin{equation}
\suinter{1}{2}{3}{4}{l}{m} = \sum_{m,m'}
C^{j_1 j_2 l}_{m_1 m_2 m}
C^{j_3 j_4 l'}_{m_3 m_4 m'}
C^{l l' 0}_{m m' 0} \,.
\end{equation}
We denote the normalisation of the intertwiners with $n$: 
\begin{align}
\sum_{\{ m \}}
\suinter{1}{2}{3}{4}{l}{m}
\suinter{1}{2}{3}{4}{l'}{m}
= \delta^{l,l'} n(j_1,j_2,j_3, j_4,l)\, ,
\end{align}
but the specific details of the normalisation do not matter, at this stage.

\subsection{Dirac deltas for left and right gauge invariance}
For our calculations we need the Dirac delta on the homogeneous space obtained imposing gauge invariance on the right and on the left. We report their Peter--Weyl decomposition for completeness.
For the right gauge invariant delta we have:
\begin{align}
\int d\gamma \prod_{i=1}^{4}\delta(g_{\va (v,i)}\gamma g_{\va (v',i)}^{-1})
&= \sum_{\{ j \}} d_{j_1}d_{j_2}d_{j_3}d_{j_4} 
\prod_{i=1}^{4} \delta^{m_i n_i}
D^{j_i}_{m_i s_i}(g_{\va (v,i)})
\suinter{1}{2}{3}{4}{l}{s}
\suinter{1}{2}{3}{4}{l}{r}
D^{j_i}_{r_i n_i}(g_{\va (v',i)}^{-1}) = \nonumber
\\
&=\sum_{\{ j \}} d_{j_1}d_{j_2}d_{j_3}d_{j_4} 
\prod_{i=1}^{4} 
D^{j_i}_{r_i s_i}(g_{\va (v',i)}^{-1}g_{\va (v,i)})
\suinter{1}{2}{3}{4}{l}{s}
\suinter{1}{2}{3}{4}{l}{r}
=
\nonumber
\\
&= \Delta_R(g_v,g_{v'})\,. 
\end{align}
For the left gauge invariant delta:
\begin{align}
\int d\gamma \prod_{i=1}^{4}\delta(\gamma g_{\va (v,i)} g_{\va (v',i)}^{-1})
&= \sum_{\{ j \}} d_{j_1}d_{j_2}d_{j_3}d_{j_4} 
\int \prod_{i=1}^{4} \delta^{m_i n_i}
D^{j_i}_{m_i s_i} (\gamma) d\gamma
D^{j_i}_{s_in_i}(g_{\va (v,i)}g_{\va (v',i)}^{-1}) =
\nonumber
\\
& =
\sum_{\{ j,l \}} d_{j_1}d_{j_2}d_{j_3}d_{j_4} 
\suinter{1}{2}{3}{4}{l}{m}
\suinter{1}{2}{3}{4}{l}{s}
\prod_{i=1}^{4}D^{j_i}_{s_im_i}(g_{\va (v,i)}g_{\va (v',i)}^{-1}) \nonumber
=\\
&= \Delta_{L}(g_{v},g_{v'}) \,.
\end{align}

\subsection{Commutation relations among operators}
We now examine the fate of the commutation relations between the new ladder operators:
\begin{equation}
\hsigma(h_v) = \int dg_v \sigma(h_vg_v) \hphi(g_v)\, ,
\qquad
\hsigmad(h_v) = \int dg_v \overline{\sigma}(h_vg_v) \hphid(g_v) \, .
\end{equation}
If we first use the commutation relations for the original ladder operators and then
the Peter--Weyl decomposition of the wave-functions we obtain:
\begin{align}
[\hsigma(h_v),\hsigmad(k_v)] &= 
\int dg_v
\sigma(h_vg_v) 
\overline{\sigma}(k_vg_v) =\nonumber
 \\&
\sum_{\{ j\}, l_{L}, l'_{L}, l_{R}}
\suinter{1}{2}{3}{4}{l_L}{m}
\suinter{1}{2}{3}{4}{l'_L}{m'}
n(j_{1}j_{2}j_{3}j_{4},l_R)
\sigma^{j_{1}j_{2}j_{3}j_{4}l_{L}l_{R}}
\overline{
\sigma^{j_{1}j_{2}j_{3}j_{4}l'_{L}l_{R}}}
\prod_{i=1}^{4} \frac{1}{d_{j_{i}}} D^{j_{i}}_{m_{i}m_{i}'}(h_{\va (v,i)}k_{\va (v,i)}^{-1})\,.
\end{align}

Therefore, if we take
\begin{align}
\sigma^{j_{1}j_2 j_3 j_4 l_{L} l_{R}} = \sigma^{j_{1}j_2 j_3 j_4 l_L} \delta^{l_L l_R}\, , \\
|\sigma^{j_1 j_2 j_3 j_4 l}|^2 = \frac{ d^2_{j_1}d^2_{j_2}d^2_{j_3}d^2_{j_4}}{n(j_{1}j_2 j_3 j_4 l)} \,,
\end{align}
we have that:
\begin{equation}
[\hsigma(h_v),\hsigmad(k_v)] = \Delta_L(h_v,k_v) \,.
\end{equation}

\acknowledgments{LS has been supported by the Templeton Foundation through the grant number PS-GRAV/1401.}


\begin{thebibliography}{10}


\bibitem{LQG}  A.~Ashtekar and J.~Lewandowski,
  ``Background independent quantum gravity: A Status report,''
  Class.\ Quant.\ Grav.\  {\bf 21}, R53 (2004)
  [gr-qc/0404018].

  C. Rovelli, ``Quantum Gravity'', Cambridge, UK: Univ. Pr. (2004) 480 p.

  T.~Thiemann,
  ``Modern canonical quantum general relativity'',
  Cambridge, UK: Cambridge Univ. Pr. (2007) 819 p.

   \bibitem{SF} 
    A.~Perez,
  ``The Spin Foam Approach to Quantum Gravity,''
  Living Rev.\ Rel.\  {\bf 16}, 3 (2013)
  [arXiv:1205.2019 [gr-qc]].

C.~Rovelli,
  ``Zakopane lectures on loop gravity,''
  PoS QGQGS {\bf 2011}, 003 (2011)
  [arXiv:1102.3660 [gr-qc]].

   \bibitem{GFT}  
 D.~Oriti, ``The Group field theory approach to quantum gravity,''
  in {\sl Approaches to quantum gravity},  D. Oriti (ed.) 
(Cambridge University Press, Cambridge UK, 2009),
  [gr-qc/0607032].

 L.~Freidel,
  ``Group field theory: An Overview,''
  Int.\ J.\ Theor.\ Phys.\  {\bf 44}, 1769 (2005)
  [hep-th/0505016].


  D.~Oriti, ``Quantum Gravity as a quantum field theory of simplicial geometry,''
 in {\sl Mathematical and Physical Aspects of Quantum Gravity,} 
  B. Fauser, et al. (eds) (Birkhaeuser, Basel, 2006), [gr-qc/0512103]; 

 A.~Baratin and D.~Oriti,
  ``Ten questions on Group Field Theory (and their tentative answers),''
  J.\ Phys.\ Conf.\ Ser.\  {\bf 360}, 012002 (2012)
  [arXiv:1112.3270 [gr-qc]].

T.~Krajewski,
 ``Group field theories,''
  PoS QGQGS {\bf 2011}, 005 (2011)
  [arXiv:1210.6257 [gr-qc]].

D.~Oriti,
  ``The Group field theory approach to quantum gravity: Some recent results,''
in {\sl The Planck Scale},  J. Kowalski-Glikman, et al. (eds) 
AIP: conference proceedings (2009),  arXiv:0912.2441 [hep-th]. 

\bibitem{OritiMicroDyn}
  D.~Oriti,
  ``The microscopic dynamics of quantum space as a group field theory,''
  in: ``Foundations of space and time'', G. Ellis, J. Marugan, A. Weltman (eds.), Cambridge University Press (2012), arXiv:1110.5606 [hep-th]     

   \bibitem{TM}
  R.~Gurau and J.~P.~Ryan,
  ``Colored Tensor Models - a review,''
  SIGMA {\bf 8} (2012) 020
  [arXiv:1109.4812 [hep-th]].

 \bibitem{LQC}  A.~Ashtekar and P.~Singh,
  ``Loop Quantum Cosmology: A Status Report,''
  Class.\ Quant.\ Grav.\  {\bf 28}, 213001 (2011)
  [arXiv:1108.0893 [gr-qc]].

 K.~Banerjee, G.~Calcagni and M.~Martin-Benito,
  ``Introduction to loop quantum cosmology,''
  SIGMA {\bf 8}, 016 (2012)
  [arXiv:1109.6801 [gr-qc]].
 
 
  E.~Wilson-Ewing,
  ``Lattice loop quantum cosmology: scalar perturbations,''
  Class.\ Quant.\ Grav.\  {\bf 29}, 215013 (2012)
  [arXiv:1205.3370 [gr-qc]].

 \bibitem{ABCK}
  A.~Ashtekar, J.~Baez, A.~Corichi and K.~Krasnov,
  ``Quantum geometry and black hole entropy,''
  Phys.\ Rev.\ Lett.\  {\bf 80} (1998) 904.
  [arXiv:gr-qc/9710007].  
  
   A.~Ashtekar, J.C.~Baez and K.~Krasnov,
 ``Quantum geometry of isolated horizons and black hole entropy,''
  Adv.\ Theor.\ Math.\ Phys.\  {\bf 4} (2000) 1.
  [arXiv:gr-qc/0005126].
  
   M.~Domagala and J.~Lewandowski,
  ``Black hole entropy from quantum geometry,''
  Class.\ Quant.\ Grav.\  {\bf 21} (2004) 5233
  [gr-qc/0407051].

\bibitem{ENPP} 
  J.~Engle, K.~Noui, A.~Perez and D.~Pranzetti,
  ``Black hole entropy from an SU(2)-invariant formulation of Type I isolated horizons,''
  Phys.\ Rev.\ D {\bf 82}, 044050 (2010)
  [arXiv:1006.0634 [gr-qc]].

  J.~Engle, K.~Noui, A.~Perez, D.~Pranzetti,
  ``The SU(2) Black Hole entropy revisited,''
  JHEP {\bf 1105}, 016 (2011).
  [arXiv:1103.2723 [gr-qc]].
  
  A.~Ghosh and A.~Perez,
  ``Black hole entropy and isolated horizons thermodynamics,''
  Phys.\ Rev.\ Lett.\  {\bf 107} (2011) 241301
   [Erratum-ibid.\  {\bf 108} (2012) 169901]
  [arXiv:1107.1320 [gr-qc]].  
  
   J.~Diaz-Polo and D.~Pranzetti,
  ``Isolated Horizons and Black Hole Entropy In Loop Quantum Gravity,''
  SIGMA {\bf 8}, 048 (2012)
  [arXiv:1112.0291 [gr-qc]].
  
   E.~Frodden, M.~Geiller, K.~Noui and A.~Perez,
  ``Black Hole Entropy from complex Ashtekar variables,''
  Europhys.\ Lett.\  {\bf 107}, 10005 (2014)
  [arXiv:1212.4060 [gr-qc]].
  
   A.~Ghosh and D.~Pranzetti,
  ``CFT/Gravity Correspondence on the Isolated Horizon,''
  Nucl.\ Phys.\ B {\bf 889}, 1 (2014)
  [arXiv:1405.7056 [gr-qc]].
  
 D.~Pranzetti and H.~Sahlmann,
  ``Horizon entropy with loop quantum gravity methods,''
  Phys.\ Lett.\ B {\bf 746}, 209 (2015)
  [arXiv:1412.7435 [gr-qc]].
  
  \bibitem{Pranzetti}
  D.~Pranzetti,
  ``Radiation from quantum weakly dynamical horizons in LQG,''
  Phys.\ Rev.\ Lett.\  {\bf 109} (2012) 011301
  [arXiv:1204.0702 [gr-qc]].
  
   D.~Pranzetti,
  ``Dynamical evaporation of quantum horizons,''
  Class.\ Quant.\ Grav.\  {\bf 30}, 165004 (2013)
  [arXiv:1211.2702 [gr-qc]].

\bibitem{biancarenorm}  B.~Dittrich, F.~C.~Eckert and M.~Martin-Benito,
  ``Coarse graining methods for spin net and spin foam models,''
  New J.\ Phys.\  {\bf 14}, 035008 (2012)
  [arXiv:1109.4927 [gr-qc]].

B.~Bahr, B.~Dittrich, F.~Hellmann and W.~Kaminski,
  ``Holonomy Spin Foam Models: Definition and Coarse Graining,''
  Phys.\ Rev.\ D {\bf 87}, 044048 (2013)
  [arXiv:1208.3388 [gr-qc]].

 B.~Dittrich, M.~Mart\`in-Benito and E.~Schnetter,
  ``Coarse graining of spin net models: dynamics of intertwiners,''
  New J.\ Phys.\  {\bf 15}, 103004 (2013)
  [arXiv:1306.2987 [gr-qc]].

B.~Dittrich,  S.~Mizera, S.~Steinhaus, \lq\lq Decorated tensor network renormalization for lattice gauge theories and spin foam models", arXiv:1409.2407 [gr-qc]

B.~Bahr, \lq\lq On backgroung-independent renormalisation of spin foam models", arXiv:1407.7746 [gr-qc] 
  
 \bibitem{GFTrenorm}   J.~Ben Geloun and V.~Rivasseau,
  ``A Renormalizable 4-Dimensional Tensor Field Theory,''
  Commun.\ Math.\ Phys.\  {\bf 318}, 69 (2013)
  [arXiv:1111.4997 [hep-th]].
  %%CITATION = ARXIV:1111.4997;%%

  S.~Carrozza, D.~Oriti and V.~Rivasseau,
  ``Renormalization of Tensorial Group Field Theories: Abelian U(1) Models in Four Dimensions,''
  Commun.\ Math.\ Phys.\  {\bf 327}, 603 (2014)
  [arXiv:1207.6734 [hep-th]].
  %%CITATION = ARXIV:1207.6734;%%
 
  D.~O.~Samary and F.~Vignes-Tourneret,
  ``Just Renormalizable TGFT's on $U(1)^{d}$ with Gauge Invariance,''
  Commun.\ Math.\ Phys.\  {\bf 329}, 545 (2014)
  [arXiv:1211.2618 [hep-th]].
  %%CITATION = ARXIV:1211.2618;%%

  S.~Carrozza, D.~Oriti and V.~Rivasseau,
  ``Renormalization of a SU(2) Tensorial Group Field Theory in Three Dimensions,''
  Commun.\ Math.\ Phys.\  {\bf 330}, 581 (2014)
  [arXiv:1303.6772 [hep-th]].
  %%CITATION = ARXIV:1303.6772;%%


  J.~Ben Geloun,
  ``Renormalizable Models in Rank $d\geq 2$ Tensorial Group Field Theory,'' 
Commun. Math. Phys. {\bf 332}, 117--188 (2014)
  [arXiv:1306.1201 [hep-th]].
  %%CITATION = ARXIV:1306.1201;%%

  J.~Ben Geloun,
  ``On the finite amplitudes for open graphs in Abelian dynamical colored Boulatov-Ooguri models,''
  J.\ Phys.\ A {\bf 46}, 402002 (2013)
  [arXiv:1307.8299 [hep-th]].
  %%CITATION = ARXIV:1307.8299;%%


%\cite{Carrozza:2013mna}
  S.~Carrozza,
  ``Tensorial methods and renormalization in Group Field Theories,''
Springer Theses, 2014 (Springer, NY, 2014),  arXiv:1310.3736 [hep-th].
  %%CITATION = ARXIV:1310.3736;%%

  J.~Ben Geloun,
  ``Two and four-loop $\beta$-functions of rank 4 renormalizable tensor field theories,''
  Class.\ Quant.\ Grav.\  {\bf 29}, 235011 (2012)
  [arXiv:1205.5513 [hep-th]].
  %%CITATION = ARXIV:1205.5513;%%

  D.~O.~Samary,
  ``Beta functions of $U(1)^d$ gauge invariant just renormalizable tensor models,''
  Phys.\ Rev.\ D {\bf 88}, 105003 (2013)
  [arXiv:1303.7256 [hep-th]].
  %%CITATION = ARXIV:1303.7256;%%

  S.~Carrozza,
  ``Discrete Renormalization Group for SU(2) Tensorial Group Field Theory,''
  arXiv:1407.4615 [hep-th].
  %%CITATION = ARXIV:1407.4615;%%

   
 \bibitem{coherent states} T.~Thiemann,
  ``Gauge field theory coherent states (GCS): 1. General properties,''
  Class.\ Quant.\ Grav.\  {\bf 18}, 2025 (2001)
  [hep-th/0005233].
  
   T.~Thiemann and O.~Winkler,
  ``Gauge field theory coherent states (GCS). 2. Peakedness properties,''
  Class.\ Quant.\ Grav.\  {\bf 18}, 2561 (2001)
  [hep-th/0005237].
  
    T.~Thiemann and O.~Winkler,
  ``Gauge field theory coherent states (GCS): 3. Ehrenfest theorems,''
  Class.\ Quant.\ Grav.\  {\bf 18}, 4629 (2001)
  [hep-th/0005234].
  
    T.~Thiemann and O.~Winkler,
  ``Gauge field theory coherent states (GCS) 4: Infinite tensor product and thermodynamical limit,''
  Class.\ Quant.\ Grav.\  {\bf 18}, 4997 (2001)
  [hep-th/0005235].
  
 \bibitem{OPS} 
  D.~Oriti, R.~Pereira and L.~Sindoni,
  Coherent states in quantum gravity: a construction based on the flux representation of LQG,
  {\em J.\ Phys.\ A} {\bf 45} (2012) 244004,
  arXiv:1110.5885[gr-qc];  A.~Pittelli and L.~Sindoni,
  New coherent states and modified heat equations,
  arXiv:1301.3113[gr-qc];
   D. Oriti, R. Pereira, and L. Sindoni, Coherent states for quantum gravity: toward collective variables, {\em Class. Quant. Grav.} {\bf 29} (2012), 135002, arXiv:1202.0526[gr-qc].
 
  
\bibitem{asymptSF} 
M.~X.~Han and M.~Zhang,
 ``Asymptotics of Spinfoam Amplitude on Simplicial Manifold: Euclidean Theory,''
  Class.\ Quant.\ Grav.\  {\bf 29}, 165004 (2012) [arXiv:1109.0500 [gr-qc]].


\bibitem{AlesciCianfrani} 
  E.~Alesci and F.~Cianfrani,
  ``A new perspective on cosmology in Loop Quantum Gravity,''
  Europhys.\ Lett.\  {\bf 104}, 10001 (2013)
  [arXiv:1210.4504 [gr-qc]].
  
   E.~Alesci and F.~Cianfrani,
  ``Quantum-Reduced Loop Gravity: Cosmology,''
  Phys.\ Rev.\ D {\bf 87}, no. 8, 083521 (2013)
  [arXiv:1301.2245 [gr-qc]].
  
  N. Bodendorfer,  \lq\lq A quantum reduction to Bianchi I models in loop quantum gravity", arXiv:1410.5608 [gr-qc]

\bibitem{SFcosmology} 
  E.~Bianchi, C.~Rovelli and F.~Vidotto,
  ``Towards Spinfoam Cosmology,''
  Phys.\ Rev.\ D {\bf 82}, 084035 (2010)
  [arXiv:1003.3483 [gr-qc]].


 \bibitem{cosmoshort}
  S.~Gielen, D.~Oriti and L.~Sindoni,
  ``Cosmology from Group Field Theory Formalism for Quantum Gravity,''
  Phys.\ Rev.\ Lett.\  {\bf 111} (2013) 031301
  [arXiv:1303.3576 [gr-qc]].
 
 
 \bibitem{cosmolong}
 S.~Gielen, D.~Oriti and L.~Sindoni,
  ``Homogeneous cosmologies as group field theory condensates,''
  JHEP {\bf 1406}, 013 (2014)
  arXiv:1311.1238 [gr-qc].
  
  \bibitem{Gielen}
  S.~Gielen,
  ``Quantum cosmology of (loop) quantum gravity condensates: An example,''
  arXiv:1404.2944 [gr-qc]; G. Calcagni, Loop quantum cosmology from group field theory, 
{\em Phys.\ Rev.\ D} {\bf 90} (2014) 064047, arXiv:1407.8166[gr-qc].

  
  \bibitem{GO} 
  S.~Gielen and D.~Oriti,
  ``Quantum cosmology from quantum gravity condensates: cosmological variables and lattice-refined dynamics,''
  New J.\ Phys.\  {\bf 16}, 123004 (2014)
  [arXiv:1407.8167 [gr-qc]]; S.~Gielen,
  ``Perturbing a quantum gravity condensate,''
  arXiv:1411.1077 [gr-qc]. 


  \bibitem{fidelity}
   L.~Sindoni,
  ``Effective equations for GFT condensates from fidelity,''
  arXiv:1408.3095 [gr-qc].


 \bibitem{GFT-LQG} 
  D.~Oriti,
  ``Group field theory as the 2nd quantization of Loop Quantum Gravity,''
  arXiv:1310.7786 [gr-qc].
  
 D.~Oriti,
  ``Group Field Theory and Loop Quantum Gravity,''
  arXiv:1408.7112 [gr-qc].
 
 D.~Oriti, J.~P.~Ryan and J.~Thürigen,
  ``Group field theories for all loop quantum gravity,''
  New J.\ Phys.\  {\bf 17}, no. 2, 023042 (2015)
    arXiv:1409.3150 [gr-qc].
  
  \bibitem{matrixmodels}
  P.~Di Francesco, P.~H.~Ginsparg and J.~Zinn-Justin,
  ``2-D Gravity and random matrices,''
  Phys.\ Rept.\  {\bf 254} (1995) 1
  [hep-th/9306153].

  
  

\bibitem{GFTnoncomm}
  A.~Baratin and D.~Oriti,
  ``Group field theory with non-commutative metric variables,''
  Phys.\ Rev.\ Lett.\  {\bf 105} (2010) 221302
  [arXiv:1002.4723 [hep-th]].  
  
\bibitem{Baratin}
  A.~Baratin, B.~Dittrich, D.~Oriti and J.~Tambornino,
  ``Non-commutative flux representation for loop quantum gravity,''
  Class.\ Quant.\ Grav.\  {\bf 28} (2011) 175011
  [arXiv:1004.3450 [hep-th]].    

\bibitem{twistor}
L.~Freidel and S.~Speziale,
  ``From twistors to twisted geometries,''
  Phys.\ Rev.\ D {\bf 82} (2010) 084041
  [arXiv:1006.0199 [gr-qc]].
  
W.~M.~Wieland,
  ``Twistorial phase space for complex Ashtekar variables,''
  Class.\ Quant.\ Grav.\  {\bf 29} (2012) 045007
  [arXiv:1107.5002 [gr-qc]].
  
E.~R.~Livine, S.~Speziale and J.~Tambornino,
  ``Twistor Networks and Covariant Twisted Geometries,''
  Phys.\ Rev.\ D {\bf 85} (2012) 064002
  [arXiv:1108.0369 [gr-qc]].
  
 S.~Speziale and W.~M.~Wieland,
  ``The twistorial structure of loop-gravity transition amplitudes,''
  Phys.\ Rev.\ D {\bf 86} (2012) 124023
  [arXiv:1207.6348 [gr-qc]].

\bibitem{Hu} B.~L.~Hu,
  ``Can space-time be a condensate?,''
  Int.\ J.\ Theor.\ Phys.\  {\bf 44}, 1785 (2005)
  [gr-qc/0503067].
    
 \bibitem{danieleemergence}
D.~Oriti,
  ``Group field theory as the microscopic description of the quantum spacetime fluid: A New perspective on the continuum in quantum gravity,''
  PoS QG {\bf -PH} (2007) 030
  [arXiv:0710.3276 [gr-qc]].

 D.~Oriti,
  ``Disappearance and emergence of space and time in quantum gravity,''
  Stud.\ Hist.\ Philos.\ Mod.\ Phys.\  {\bf 46} (2014) 186
  [arXiv:1302.2849 [physics.hist-ph]].
   
    \bibitem{FRG-GFT}
T.~Krajewski and R.~Toriumi,
  ``Polchinski's equation for group field theory,''
  Fortsch.\ Phys.\  {\bf 62}, 855 (2014).
  
  D. Benedetti, J. Ben Geloun, D. Oriti, ``Functional Renormalisation Group Approach for Tensorial Group Field Theory: a Rank-3 Model,'' arXiv:1411.3180 [hep-th].


    \bibitem{Tim} T.~A.~Koslowski,
  ``Dynamical Quantum Geometry (DQG Programme),''
  arXiv:0709.3465 [gr-qc].
    
    \bibitem{Hanno-Tim} 
  T.~Koslowski and H.~Sahlmann,
  ``Loop quantum gravity vacuum with nondegenerate geometry,''
  SIGMA {\bf 8}, 026 (2012)
  [arXiv:1109.4688 [gr-qc]].
    
\bibitem{leggett}   A.~Leggett, ``Quantum Liquids,''
Oxford University Press (2006). 

 A.~J.~Leggett,
  ``Bose-Einstein condensation in the alkali gases: Some fundamental concepts,''
  Rev.\ Mod.\ Phys.\  {\bf 73}, 307 (2001).

    
  \bibitem{Flower} 
  L.~Freidel and E.~R.~Livine,
  ``Spin networks for noncompact groups,''
  J.\ Math.\ Phys.\  {\bf 44}, 1322 (2003)
  [hep-th/0205268].


  \bibitem{eteradanny}
 E.~R.~Livine and D.~R.~Terno,
  ``Quantum black holes: Entropy and entanglement on the horizon,''
  Nucl.\ Phys.\ B {\bf 741} (2006) 131
  [gr-qc/0508085].
  
  \bibitem{martin} 
  M.~Bojowald and H.~A.~Kastrup,
  ``Quantum symmetry reduction for diffeomorphism invariant theories of connections,''
  Class.\ Quant.\ Grav.\  {\bf 17}, 3009 (2000)
  [hep-th/9907042].
  
   M.~Bojowald and R.~Swiderski,
  ``Spherically symmetric quantum geometry: Hamiltonian constraint,''
  Class.\ Quant.\ Grav.\  {\bf 23}, 2129 (2006)
  [gr-qc/0511108].

  \bibitem{IH} A.~Ashtekar, C.~Beetle and S.~Fairhurst,
  ``Mechanics of isolated horizons,''
  Class.\ Quant.\ Grav.\  {\bf 17}, 253 (2000)
  [gr-qc/9907068].

\bibitem{KMS} 
  D.~Pranzetti,
  ``Geometric temperature and entropy of quantum isolated horizons,''
  Phys.\ Rev.\ D {\bf 89}, 104046 (2014)
  [arXiv:1305.6714 [gr-qc]].
  
  \bibitem{BH-condensate} 
  D.~Oriti, D.~Pranzetti and L.~Sindoni,
  ``Entropy of isolated horizons from quantum gravity condensates,''
  arXiv:1510.06991 [gr-qc].



\end{thebibliography}
\end{document}